\documentclass[12pt]{article}

\usepackage{graphicx}
\usepackage{amsmath,amscd,cite}

\begin{document}

\begin{titlepage}
\title{Solution of the Crow-Kimura and Eigen models for
alphabets of arbitrary size
by Schwinger spin coherent states}

\author{
Enrique Mu\~noz, Jeong-Man Park, and Michael W. Deem\\
~\\
Department of Physics \& Astronomy\\
Rice University\\
6100 Main St., Houston, TX 77005--1892
}
\maketitle

\bigskip

\end{titlepage}

\begin{abstract}
To represent the evolution of nucleic acid and protein sequence,
we express the  
parallel and Eigen models for molecular evolution in terms of a functional
integral representation with an $h$-letter alphabet, lifting the 
two-state, purine/pyrimidine assumption often  made in quasi-species theory. 
For arbitrary $h$ and a general mutation scheme,
we obtain the solution of this model in terms of a maximum
principle.  
Euler's theorem for homogeneous functions is used to derive
this `thermodynamic' formulation of evolution.
The general result for the parallel model reduces
to known results for the purine/pyrimidine $h=2$ alphabet and the
nucleic acid $h=4$ alphabet for 
the Kimura 3 ST mutation scheme. Examples are presented
for the $h=4$ and $h=20$ cases.
We derive the maximum principle for the Eigen model for general $h$.
The general result for the Eigen model reduces to a known
result for $h=2$.
Examples are presented for the nucleic acid $h=4$ and
the amino acid $h=20$ alphabet. An error catastrophe 
phase transition occurs in these
models, and 
the order of the  phase transition changes from
second to first order for smooth fitness functions when the alphabet
size is increased beyond two letters to the generic case.
As examples, we analyze the general analytic
solution for sharp peak, linear, quadratic, and quartic fitness functions.  
\end{abstract}

\section{Introduction}

There are two classical physical models
of molecular evolution: the Eigen model \cite{Eigen71,Eigen88,Eigen89} 
and the Parallel or
Crow-Kimura model \cite{Kimura70}. These models were originally formulated
in the language of chemical kinetics \cite{Eigen71}, by a large system of
differential equations representing the time evolution
of the relative frequencies of each sequence type. Quasi-species
models capture the basic microscopic processes of
mutation and replication, for an infinite population of
binary sequences. The most remarkable feature of these models
is the existence of a phase transition, termed the "error threshold"
\cite{Eigen71,Biebricher05}, 
when the mutation rate is below a critical value, separating
a disordered non-selective phase from an organized or
"quasi-species" phase. The quasi-species is characterized
by a population of closely related mutants, rather than by
identical sequences \cite{Eigen71,Eigen88,Eigen89}, and its
emergence is related to the auto-catalytic character of the
replication process \cite{Eigen71,Biebricher05}, 
which exponentially enriches the proportion of the fittest mutants
in the population. Experimental studies provide support for 
quasi-species theory in the evolution of RNA viruses 
\cite{Domingo78,Domingo05}.

The choice of a binary alphabet, which simplifies
the mathematical and numerical analysis of the theory, represents
a coarse graining of the four-letter alphabet of
the nucleic acids DNA/RNA (A,C,G,T/U), 
by considering the two basic chemical structures
of nitrogenated bases, purines (A,G) and pyrimidines (C,T/U).
The choice of a four-letter alphabet represents nucleic
acids. A 20-letter alphabet represents amino acids and
protein structure and permits a close connection between
sequence and fitness.

Most numerical and analytical studies on quasi-species
models consider the binary alphabet simplification 
\cite{Eigen71,Eigen88,Eigen89,Kimura70}. In particular, 
the assumption of a binary
alphabet allows for an exact mapping of the quasi-species
models into a 2D Ising model 
\cite{Leuthausser87,Tarazona92}, or into a quantum spin chain
\cite{Baake97,Baake98,Baake01,Saakian04b,Saakian06}. 
An 
exception is \cite{Hermisson01}, where a four-letter alphabet
was studied by a quantum spin chain representation of the
parallel model. Other approaches to the nucleic acid
evolution problem have been presented in \cite{Garske03,Garske04}.
In all these studies it has been shown, through the application
of different methodologies, that the steady state mean fitness of
the population can be expressed in terms of a maximum principle, in
the limit of infinite sequence length ($N\rightarrow\infty$). 
The Frobenius-Perrone theorem guarantees that there is a unique
steady-state population distribution.
It has been shown \cite{Baake05}
that for a general family of linear (or effectively linear) models
that evolve according to a matrix $H = M + R$, with $M$ a Markov
generator (typically representing mutations) and $R$ a diagonal matrix 
(usually representing replication or degradation) of dimension
$N\rightarrow\infty$, the Frobenius-Perrone 
largest eigenvalue can be expressed in terms of a Raleigh-Ritz variational
problem. The high dimensional variational problem can be
reduced to a low-dimensional maximum principle with an error $O(1/N)$, 
when basic symmetries
can be assumed in the evolution matrix, such as permutation invariance
of the replication rate, or symmetric mutation rates, 
which allows for a lumping \cite{Baake05} 
of the large sequence space into sequence types or classes. This
analysis was applied in \cite{Garske04} to
obtain a variational expression for the mean fitness in
the Kimura model with a four letters alphabet. We note
that the Eigen model, where replication and (multiple) mutations 
are correlated, possesses a different algebraic structure than the 
general family of models studied in \cite{Baake05}. In the Eigen model,
the evolution matrix is of the form $H = Q \times R$, with $Q$ representing
the mutation matrix, and $R$ a diagonal replication matrix. 

In this article, we present exact analytical solutions
of the $h$-alphabet Crow-Kimura and Eigen models by means of
a quantum field theory. Our method generalizes the Schwinger
spin coherent field theory for the binary alphabet 
in \cite{Park06} to an
alphabet of arbitrary size $h$. 
This method has also been recently applied in the solution
of a model that includes transfer of genetic material between
sequences in quasi-species theory \cite{Park07,Munoz08}, and
two-parent recombination \cite{Munoz08}.
For the parallel model, we present exact analytical solutions of this
field theory, in terms of a maximum principle, for the steady state
mean fitness of the population and average composition, Eq.\ 
(\ref{eqGePar20gen_1}).
We present as examples,
results for the Kimura 3 ST mutation 
scheme \cite{Kimura81}, Eq.\ (54). We develop in detail the 
result for the symmetric mutation rate scheme, Eq.\ (55), for
four different examples of microscopic fitness functions: sharp
peak Eqs.\ (57) and (58), Fujiyama landscape Eqs.\ (60)--(63), 
a quadratic landscape Eqs.\ (65)--(69), and a quartic landscape
Eqs.\ (71) and (72). 
In section 2.9, we apply our general formula to derive
the mean fitness for the symmetric, general $h$ case
and discuss the $h=20$ amino acid alphabet, Eq.\ (74).
For the symmetric case, we present results for the sharp
peak, Eqs.\ (75)--(77), and the quadratic case, Eq.\ (77).

For the Eigen model, we present the exact expression for arbitrary
alphabet size $h$, Eq.\ (106). As an example, we apply the general
expression to 
a mutation scheme analogous to the Kimura 3 ST \cite{Kimura81}, 
Eq.\ (112). We analyze
in detail the solution for the symmetric mutations rate, Eq.\ (113), for
four different examples of
microscopic fitness functions: sharp peak Eqs.\ (114) and (115), 
Fujiyama landscape Eqs.\ (116--119), quadratic fitness landscape
Eqs.\ (121--123), and quartic fitness landscape  Eqs.\ (125) and (126). 
In section 3.8, we apply our general solution to derive
the mean fitness for the symmetric, general $h$ case and discuss the
$h=20$ amino acid alphabet, Eq.\ (127).
For the symmetric case, we present results for the sharp
peak, Eqs.\ (128)--(129), and the quadratic case.

These results bring quasi-species theory closer to the real
microscopic evolutionary dynamics that occurs in the natural
four-letter alphabet of nucleic acids or
the 20-letter alphabet of amino acids.

\section{The parallel model for an alphabet of size $h$}
The parallel model \cite{Kimura70} describes the continuous time evolution of
an infinite size population of viral genetic sequences. The 
evolutionary dynamics is driven by point mutations and 
selection, with mutations occurring in parallel and independently
of viral replication. Each viral genome is represented as a 
sequence of N letters, from an alphabet of size $h$, and therefore
the total number of different viral genomes in the population is
$h^{N}$. If we describe a viral genetic sequence in the alphabet of nucleic
acids (DNA or RNA), the natural choice would be $h=4$, and explicitly
the alphabet corresponds to (A,C,G,T or U). It is common,
to simplify the theoretical analysis, to choose instead a coarse grained
alphabet of size $h = 2$, by 'lumping' together purines (A, T or U)
and pyrimidines (C,G). Alternatively, to describe
evolution at the scale of protein sequences, the natural choice is 
to consider the $h=20$ amino acid alphabet. 
We here consider the case of general $h$.

The probability $p_i$ for a virus to have a genetic sequence
$S_i$, $1 \leq i \leq h^{N}$, evolves according to the following
system of nonlinear differential equations
\begin{eqnarray}
\frac{dp_{i}}{dt} = p_{i}(r_{i} - \sum_{j=1}^{h^{N}}r_{j}p_{j})
+\sum_{j=1}^{h^{N}}\mu_{ij}p_{j}
\label{eq1}
\end{eqnarray}  
Here $r_{i}$ is replication rate of sequence $S_i$, 
and $\mu_{ij}$ is the mutation rate 
from sequence $S_{j}$ into sequence $S_{i}$.
The nonlinear term in Eq.\ (\ref{eq1}) represents the average replication
rate in the population, or mean fitness.
This non-linear term 
enforces the conservation of probability, $\sum_{i}p_{i}=1$. This
term
can be removed through a simple exponential transformation, to
obtain the linear system of differential equations
\begin{eqnarray}
\frac{dq_{i}}{dt}=r_{i}q_{i} + \sum_{j=1}^{h^{N}}\mu_{ij}q_{j}
\label{eq2}
\end{eqnarray}
where $p_{i}(t)=q_{i}(t)/\sum_{j}q_{j}(t)$.

\subsection{Fitness landscape}

We will assume that the replication rate (fitness) of an individual
in the population depends on its relative composition with respect
to a wild-type sequence $S_w$.  We define the relative composition
variables, $x^\alpha$, to be the
number of letters of type $\alpha$, divided by $N$.
The number of different letters is $h$, and the
set of labels $\alpha$ refers to the set of chemical
possibilities, such as \{purine,pyrimidine\}, \{A,C,G,T\},
or the 20 amino acids.
For an alphabet of size $h$, at each
site along the sequence, there are $h-1$ independent compositions
$0\leq x^{\alpha}\leq 1$, for $1\leq \alpha \leq h-1$. 
Alternatively, these compositions may be interpreted as normalized
Hamming distances from a reference wild-type sequence.
Therefore, the replication rate for a 
sequence $S_{i}$ in the parallel
model Eq.\ (\ref{eq1}) is defined by the fitness function
\begin{eqnarray}
r_{i} = N f(x^{1},x^{2},\ldots,x^{h-1})
\label{eq6}
\end{eqnarray}
where the $x^{\alpha}$ are defined within the simplex 
$\sum_{\alpha = 1}^{h-1}x^{\alpha}\leq 1$.

\subsection{Schwinger spin coherent states representation of 
the parallel model}

We can express the parallel model in operator form, by generalizing
the method presented in \cite{Park06}. We define $h$
kinds of creation and annihilation operators: $\hat{a}^{\dagger}_{\alpha}(j),
\hat{a}_{\alpha}(j)$, $1 \leq \alpha \leq h$ and
$1\leq j\leq N$. These operators satisfy the commutation relations 
\begin{eqnarray}
\left[ \hat{a}_{\alpha}(i),\hat{a}_{\beta}^{\dagger}(j)\right] =
\delta_{\alpha\beta}\delta_{ij} \nonumber
\\
\left[ \hat{a}_{\alpha}(i),\hat{a}_{\beta}(j)\right] = 
\left[ \hat{a}_{\alpha}^{\dagger}(i),\hat{a}_{\beta}^{\dagger}(j)\right] 
= 0
\label{eq9}
\end{eqnarray}
These operators create/annihilate a sequence
letter state $1\leq \alpha \leq h$, 
at position
$1\leq j\leq N$ in the sequence. 
Since at each site there is a single letter,
we enforce the constraint
\begin{eqnarray}
\sum_{\alpha=1}^{h}\hat{a}^{\dagger}_{\alpha}(j)\hat{a}_{\alpha}(j)=1
\label{eq10}
\end{eqnarray}
for all $1\leq j \leq N$. 

We define $n_{\alpha}^{i}(j)$ as the power on 
$\hat{a}^{\dagger}_{\alpha}(j)$ for the sequence state 
$S_i$, $1\leq i \leq h^{N}$,
defined by the vectors
\begin{eqnarray}
|S_i\rangle = \prod_{j=1}^{N}|\vec{n}^{i}_{j}\rangle
\label{eq12}
\end{eqnarray}
where $|\vec{n}^{i}_{j}\rangle = 
\prod_{\alpha=1}^{h}\left[\hat{a}^{\dagger}_{\alpha}(j)
\right]^{n_{\alpha}^{i}(j)}|0\rangle$. The constraint in
Eq.\ (\ref{eq10})
ensures that the condition 
$\sum_{\alpha=1}^{h}n_{\alpha}^{i}(j)=1$ for all $i,j$.

We introduce the unnormalized population state vector 
\begin{eqnarray}
|\psi\rangle = \sum_{i=1}^{h^{N}}q(S_i)
|S_i\rangle
\label{eq13}
\end{eqnarray}
which evolves in time according to the equation 
\begin{eqnarray}
\frac{d}{dt}|\psi\rangle = - \hat{H}|\psi\rangle
\label{eq14}
\end{eqnarray}
Here, the Hamiltonian operator, to highest order in $N$, is given by
\begin{eqnarray}
-\hat{H} = \hat{m} + N f(\hat{x}^{1},\hat{x}^{2},\ldots,\hat{x}^{h-1})
\label{eq15}
\end{eqnarray}
where $\hat{m}$ represents the mutation operator, and $\hat{x}^{\alpha}$
represents the compositions in operator form.

Let us first discuss the mutation operator $\hat{m}$. In the most general
case, $h(h-1)$ possible different substitutions can occur at each
site in the sequence, i.e. $\beta \rightarrow \alpha$, with
mutation rate $\mu_{\alpha\beta}$ that need not be symmetric.

Each individual process can be written
in the operator form
\begin{eqnarray}
\vec{\hat{a}}^{\dagger}(j)\tau^{\alpha\beta}\vec{\hat{a}}(j)
=\hat{a}^{\dagger}_{\alpha}(j)\hat{a}_{\beta}(j) + 
\sum_{\gamma\ne \alpha}\hat{a}^{\dagger}_{\gamma}(j)\hat{a}_{\gamma}(j)
\label{eq16}
\end{eqnarray}
which represents the creation of letter $\alpha$ by annihilation of
letter $\beta$.
Here, the matrices $\tau^{\alpha\beta}$ are explicitly defined by
\begin{eqnarray}
\left[\tau^{\alpha\beta}\right]_{\rho\gamma} = \delta_{\rho\alpha}
\delta_{\gamma\beta} + \delta_{\rho\gamma}(1-\delta_{\rho\alpha})
\label{eq17}
\end{eqnarray}
After these definitions, the more general expression for the mutation
operator is
\begin{eqnarray}
\hat{m} = \sum_{j=1}^{N}\sum_{\alpha\ne\beta=1}^{h}\mu_{\alpha\beta}\left[
\vec{\hat{a}}^{\dagger}(j)\tau^{\alpha\beta}\vec{\hat{a}}(j) -
\vec{\hat{a}}^{\dagger}(j) \cdot \vec{\hat{a}}(j)\right]
\label{eq18}
\end{eqnarray}

Let us now consider the Schwinger spin coherent state representation of the 
average base composition terms, 

\begin{eqnarray}
\hat{x}^{\alpha} = \frac{1}{N}\sum_{j=1}^{N}\hat{a}^{\dagger}_{\alpha}(j)
\hat{a}_{\alpha}(j) \equiv \frac{1}{N}\sum_{j=1}^{N}
\vec{\hat{a}}^{\dagger}(j)\Theta^{\alpha}\vec{\hat{a}}(j)\,\,\,\,
1\leq\alpha\leq h-1
\label{eq25}
\end{eqnarray}
where we defined the matrices
\begin{eqnarray}
\left[\Theta^{\alpha}\right]_{\rho\gamma} 
= \delta_{\rho\alpha}\delta_{\gamma\alpha}
\label{eq26}
\end{eqnarray}

We introduce the vector notation
\begin{eqnarray}
\vec{\hat{x}} = (\hat{x}^{1},\hat{x}^{2},\ldots,\hat{x}^{h-1})
\label{eq27}
\end{eqnarray}
\begin{eqnarray}
{\bf{\Theta}} = (\Theta^{1},\Theta^{2},\ldots,\Theta^{h-1})
\label{eq28}
\end{eqnarray}

Considering the previous expressions, the Hamiltonian operator becomes
\begin{eqnarray}
-\hat{H} &=& Nf\left[\frac{1}{N}\sum_{j=1}^{N}\vec{\hat{a}}^{\dagger}(j)
\mathbf{\Theta}\vec{\hat{a}}(j)\right] 
+\sum_{j=1}^{N}\sum_{\alpha\ne\beta=1}^h\mu_{\alpha\beta}\left[
\vec{\hat{a}}^{\dagger}(j)\tau^{\alpha\beta}\vec{\hat{a}}(j)
-\vec{\hat{a}}^{\dagger}(j) \cdot \vec{\hat{a}}(j)\right]
\nonumber\\
\label{eq30}
\end{eqnarray}

\subsection{Functional integral representation of the parallel model}

We convert the operator representation of the parallel model into
a functional integral form by introducing Schwinger spin coherent 
states \cite{Park06}.
We define a coherent state by
\begin{eqnarray}
|\vec{z}(j)\rangle &=& e^{\vec{\hat{a}}^{\dagger}(j)\cdot\vec{z}(j)
-\vec{z}^{*}(j)\cdot\vec{\hat{a}}(j)}|0 \rangle \nonumber
\\
&=&e^{-\frac{1}{2}\vec{z}^{*}(j)\cdot\vec{z}(j)}\sum_{m_{1},m_{2},\ldots
,m_{h}=0}^{\infty}\prod_{\alpha=1}^{h}
\frac{\left[z_{\alpha}(j)\right]^{m_{\alpha}}}{\sqrt{m_{\alpha}!}}
|(m_{1},m_{2},\ldots,m_{h})_{j}\rangle\nonumber\\
\label{eq31}
\end{eqnarray}
Coherent states satisfy the completeness relation
\begin{eqnarray}
I = \int\prod_{j=1}^{N}\frac{d\vec{z}^{*}(j)d\vec{z}(j)}{\pi^{h}}
|\{\vec{z}\}\rangle\langle\{\vec{z}\}|
\label{eq32}
\end{eqnarray}
The overlap between a pair of coherent states is given by
\begin{eqnarray}
\langle\vec{z}^{\,'}(j)|\vec{z}(j)\rangle = 
e^{-\frac{1}{2}\{\vec{z}^{\,'*}(j)
\cdot[\vec{z}^{'}(j)-\vec{z}(j)]-[\vec{z}^{\,'*}(j)
-\vec{z}^{*}(j)]\cdot
\vec{z}(j)\}}
\label{eq33}
\end{eqnarray}

To enforce the constraint Eq.\ (\ref{eq10}), we introduce the projector
\begin{eqnarray}
\hat{P}=\prod_{j=1}^{N}\hat{P}(j)=\prod_{j=1}^{N}
\Delta[\vec{\hat{a}}^{\dagger}(j)\cdot\vec{\hat{a}}(j)-1]
\nonumber \\
=\int_{0}^{2\pi}\prod_{j=1}^{N}\frac{d\lambda_{j}}{2\pi}
e^{i\lambda_{j}[\vec{\hat{a}}^{\dagger}(j)\cdot\vec{\hat{a}}(j)-1]}
\label{eq34}
\end{eqnarray}

At long times, due to the Perrone-Frobenius theorem, we find
that the system evolution is dominated by the
unique largest eigenvalue, $f_{m}$, of $-\hat{H}$ and its corresponding
eigenvector $|\psi^{*}\rangle$,
such that $e^{-\hat{H}t}|\{\vec{n}_{0}\}\rangle \sim e^{f_{m}t}
|\psi^{*}\rangle$.

To evaluate this eigenvalue, we perform a Trotter factorization, for
$\epsilon = t/M$, with $M\rightarrow\infty$, and introduce resolutions
of the identity as defined by Eq.\ (\ref{eq32}) at each time slice
\cite{Park06}
\begin{eqnarray}
e^{-\hat{H}t} = \lim_{M\rightarrow\infty}\int\left[
\prod_{k=1}^{M}\prod_{j=1}^{N}
\frac{d\vec{z}^{*}_{k}(j)d\vec{z}_{k}(j)}{\pi^{h}}\right]
|\{\vec{z}_{M}\}\rangle
\prod_{k=1}^{M}\langle\{\vec{z}_{k}\}|
e^{-\epsilon\hat{H}}|\{\vec{z}_{k-1}\}\rangle
\langle\{\vec{z}_{0}\}|\nonumber\\
\label{eq35}
\end{eqnarray}

We define the partition function
\begin{eqnarray}
Z &=& {\rm{Tr}}\: e^{-\hat{H}t}\hat{P} \nonumber \\
&=& \int_{0}^{2\pi}\left[\prod_{j=1}^{N}\frac{d\lambda_{j}}{2\pi}\right]
e^{-i\lambda_{j}}
\lim_{M\rightarrow\infty} \int\left[\prod_{k=1}^{M}\prod_{j=1}^{N}
\frac{d\vec{z}^{*}_{k}(j)d\vec{z}_{k}(j)}{\pi^{h}}\right]
e^{-S[\vec{z}^{*},\vec{z}]}\nonumber\\
\label{eq36}
\end{eqnarray} 
Here, we defined
\begin{eqnarray}
e^{-S[\vec{z}^{*},\vec{z}]} = 
\prod_{k=1}^{M}\langle\{\vec{z}_{k}\}|e^{-\epsilon\hat{H}}
|\{\vec{z}_{k-1}\}\rangle
\label{eq37}
\end{eqnarray}
with the boundary condition $\vec{z}_{0}(j) = 
e^{i\lambda_{j}}\vec{z}_{M}(j)$ \cite{Park06}.
An explicit expression for the matrix element in the coherent
states representation is
\begin{eqnarray}
\langle\{\vec{z}_{k}\}|e^{-\epsilon\hat{H}}
|\{\vec{z}_{k-1}\}\rangle &=& \exp\bigg(-\frac{1}{2}\sum_{j=1}^{N}
[\vec{z}_{k}^{*}(j)\cdot\vec{z}_{k}(j)
-2\vec{z}_{k}^{*}(j)\cdot\vec{z}_{k-1}(j)+\vec{z}_{k-1}^{*}(j)
\cdot\vec{z}_{k-1}(j)]\nonumber\\
&-&\epsilon N \sum_{\alpha\ne\beta=1}^h \mu_{\alpha\beta}
+\epsilon Nf\left[\frac{1}{N}\sum_{j=1}^{N}
\vec{z}_{k}^{*}(j)\mathbf{\Theta}\vec{z}_{k-1}(j)\right]
\nonumber \\
&+&\epsilon\sum_{j=1}^{N}\vec{z}^{*}_{k}(j)\sum_{\alpha\ne\beta=1}^h
\mu_{\alpha\beta}\tau^{\alpha\beta}\vec{z}_{k-1}(j)
\bigg)\nonumber\\
\label{eq38}
\end{eqnarray}
Let us now introduce an ($h$-1)-component vector field 
$\vec{x}_{k}=(x_{k}^{1},x_{k}^{2},\ldots,x_{k}^{h-1})$, with
\begin{eqnarray}
x_{k}^{\alpha} = \frac{1}{N}\sum_{j=1}^{N}\vec{z}_{k}^{*}(j)
\Theta^{\alpha}\vec{z}_{k-1}(j)
\label{eq39}
\end{eqnarray}
We make this definition by introducing an integral representation
of the corresponding delta function
\begin{eqnarray}
1 &=& \int\mathcal{D}[\vec{x}]\prod_{k=1}^{M}\delta^{(h-1)}
\left[\vec{x}_{k}-\frac{1}{N}\sum_{j=1}^{N}
\vec{z}_{k}^{*}(j)\mathbf{\Theta}\vec{z}_{k-1}(j)\right]
\nonumber \\
&=& \int\left[\prod_{k=1}^{M}\prod_{\alpha=1}^{h-1}\frac{d\bar{x}_{k}^{\alpha}
d x_{k}^{\alpha}}{2\pi}\right]e^{i\sum_{k=1}^{M}
\vec{\bar{x}}_{k}
\cdot\vec{x}_{k}-\frac{i}{N}\sum_{k=1}^{M}\sum_{j=1}^{N}
\vec{z}^{*}_{k}(j)\vec{\bar{x}}_{k}\cdot\mathbf{\Theta}
\vec{z}_{k-1}(j) } \nonumber \\
&=& \int\left[\prod_{k=1}^{M}\prod_{\alpha=1}^{h-1}\frac{i\epsilon N 
d\bar{x}_{k}^{\alpha}
d x_{k}^{\alpha}}{2\pi}\right]e^{-\epsilon N 
\sum_{k=1}^{M}\vec{\bar{x}}_{k}
\cdot\vec{x}_{k}+\epsilon\sum_{k=1}^{M}\sum_{j=1}^{N}
\vec{z}^{*}_{k}(j)\vec{\bar{x}}_{k}\cdot\mathbf{\Theta}
\vec{z}_{k-1}(j) }\nonumber\\
\label{eq40}
\end{eqnarray}
Inserting this into the functional integral Eq.\ (\ref{eq36}), we have
\cite{Park06}
\begin{eqnarray}
Z &=& \lim_{M\rightarrow\infty}\int\mathcal{D}[\vec{\bar{x}}]
\mathcal{D}[\vec{x}]e^{\epsilon N\sum_{k=1}^{M}[
f(\vec{x}_{k})-\vec{\bar{x}}_{k}\cdot\vec{x}_{k}
-\sum_{\alpha\ne\beta=1}^{h}\mu_{\alpha\beta}]} \nonumber \\
&&\times
\int\mathcal{D}[\vec{z}^{*}]\mathcal{D}
[\vec{z}]\mathcal{D}[\lambda]\prod_{j=1}^{N}e^{-i\lambda_{j}}
e^{\epsilon\sum_{l,k=1}^{M}\vec{z}^{*}_{k}(j)S_{kl}(j)\vec{z}_{l}(j)}
\bigg | _{\{\vec{z}_{0}\}=\{e^{i\lambda_{j}}\vec{z}_{M}\}}
\label{eq41}
\end{eqnarray}
The matrix $S(j)$ has the structure
\begin{eqnarray}
S(j) = \left(\begin{array}{ccccc}
I & 0 & 0 & \ldots & -e^{-i\lambda_{j}}A_{1}(j) \\
-A_{2}(j) & I & 0 & \ldots & 0 \\
0 & -A_{3}(j) & I & \ldots & 0 \\
  &  &  & \ddots & \\
0 & & \ldots & -A_{M}(j) & I \end{array} \right)
\label{eq42}
\end{eqnarray}
where $A_{k}(j) = I + \epsilon[\sum_{\alpha\ne\beta=1}^{h}\mu_{\alpha\beta}
\tau^{\alpha\beta}+
\vec{\bar{x}}_{k}\cdot\mathbf{\Theta}]$.
After performing the Gaussian integration over the coherent state
fields, we obtain
\begin{eqnarray}
Z &=&\lim_{M\rightarrow\infty}\int\mathcal{D}[\vec{\bar{x}}]
\mathcal{D}[\vec{x}]e^{\epsilon N\sum_{k=1}^{M}[
f(\vec{x}_{k})-\vec{\bar{x}}_{k}\cdot\vec{x}_{k}
-\sum_{\alpha\ne\beta=1}^{h}\mu_{\alpha\beta}]} \nonumber \\
&&\times\int\mathcal{D}[\lambda]\prod_{j=1}^{N}e^{-i\lambda_{j}}
\left[\det S(j)\right]^{-1}
\label{eq43}
\end{eqnarray}
Here,
\begin{eqnarray}
\det S(j) &=& \det \left[ I - e^{i\lambda_{j}}\prod_{k=1}^{M}A_{k}(j)\right]
\nonumber \\
&=&\det\left[ I - e^{i\lambda_{j}}\hat{T}e^{\epsilon\sum_{k=1}^{M}
[\sum_{\alpha\ne\beta=1}^{h}\mu_{\alpha\beta}\tau^{\alpha\beta}
+\vec{\bar{x}}_{k}\cdot
\mathbf{\Theta}]}\right] \nonumber \\
&=& 
e^{{\rm{Tr}} \ln \left[ I - e^{i\lambda_{j}}\hat{T}e^{\epsilon\sum_{k=1}^{M}
[\sum_{\alpha\ne\beta=1}^{h}\mu_{\alpha\beta}\tau^{\alpha\beta}  
+\vec{\bar{x}}_{k}\cdot
\mathbf{\Theta}]}\right]}
\label{eq44}
\end{eqnarray} 
where the operator $\hat{T}$ indicates time ordering. Substituting
this result in the partition function, we obtain
\begin{eqnarray}
Z &=&  \lim_{M\rightarrow\infty}\int\mathcal{D}[\vec{\bar{x}}]
\mathcal{D}[\vec{x}]e^{\epsilon N\sum_{k=1}^{M}[
f(\vec{x}_{k})-\vec{\bar{x}}_{k}\cdot\vec{x}_{k}
-\sum_{\alpha\ne\beta=1}^{h}\mu_{\alpha\beta}]} \nonumber \\
&&\times\int\mathcal{D}[\lambda]\prod_{j=1}^{N}e^{-i\lambda_{j}}
e^{-{\rm{Tr}} \ln \left[ I - e^{i\lambda_{j}}\hat{T}e^{\epsilon\sum_{k=1}^{M}
[\sum_{\alpha\ne\beta=1}^{h}\mu_{\alpha\beta}\tau^{\alpha\beta}
+\vec{\bar{x}}_{k}\cdot\mathbf{\Theta}]}\right]} \nonumber \\
&=& \lim_{M\rightarrow\infty}\int\mathcal{D}[\vec{\bar{x}}]
\mathcal{D}[\vec{x}]e^{\epsilon N\sum_{k=1}^{M}[
f(\vec{x}_{k})-\vec{\bar{x}}_{k}\cdot\vec{x}_{k}
-\sum_{\alpha\ne\beta=1}^{h}\mu_{\alpha\beta}]}\prod_{j=1}^{N}Q
\label{eq45}
\end{eqnarray}
with
\begin{eqnarray}
Q &=&\lim_{M\rightarrow\infty}
{\rm{Tr}}\:\hat{T}\prod_{k=1}^{M}\left[ I + \epsilon(
\sum_{\alpha\ne\beta=1}^{h}\mu_{\alpha\beta}\tau^{\alpha\beta} 
+\vec{\bar{x}}_{k}\cdot
\mathbf{\Theta} )\right] \nonumber \\
&=&\lim_{M\rightarrow\infty} {\rm{Tr}}
\:\hat{T}e^{\epsilon\sum_{k=1}^{M}[\sum_{\alpha\ne\beta=1}^{h}
\mu_{\alpha\beta}\tau^{\alpha\beta}
+\vec{\bar{x}}_{k}\cdot
\mathbf{\Theta}]} \nonumber \\
&=& {\rm{Tr}}\:\hat{T}e^{\int_{0}^{t}dt'[\sum_{\alpha\ne\beta=1}^{h}
\mu_{\alpha\beta}\tau^{\alpha\beta}
+\vec{\bar{x}}(t')\cdot\mathbf{\Theta}]}
\label{eq46}
\end{eqnarray}
After taking the limit $M\rightarrow \infty$, Eq.\ (\ref{eq45}) becomes
\begin{eqnarray}
Z = \int\mathcal{D}[\vec{\bar{x}}]\mathcal{D}[\vec{x}]
e^{-S[\vec{\bar{x}},\vec{x}]}
\label{eq47}
\end{eqnarray}
where the effective action is given by
\begin{eqnarray}
S[\vec{\bar{x}},\vec{x}] &=& -N
\int_{0}^{t}dt'[f(\vec{x}(t'))-\vec{\bar{x}}(t')
\cdot\vec{x}(t')-\sum_{\alpha\ne\beta=1}^{h}\mu_{\alpha\beta}]-N\ln Q
\nonumber\\
\label{eq48}
\end{eqnarray}
Here, we have defined 
\begin{eqnarray}
Q = 
{\rm{Tr}}\:\hat{T}e^{\int_{0}^{t}dt'[\sum_{\alpha\ne\beta=1}^{h}\mu_{\alpha\beta}
\tau^{\alpha\beta}
+\sum_{\alpha=1}^{h-1}\bar{x}^{\alpha}(t')\Theta^{\alpha}]}
\label{eq57}
\end{eqnarray}

\subsection{The large N limit of the parallel model is a saddle point}

Considering that the sequence length N is very large, $N\rightarrow\infty$, 
we can evaluate the functional integral Eq.\ (\ref{eq47}) for the partition
function by looking for a saddle point. With the action defined
in Eq.\ (\ref{eq57}), we have
\begin{eqnarray}
\frac{\delta S}{\delta x^{\alpha}}\bigg | _{\vec{x}_{c},
\vec{\bar{x}}_{c}}&=&-N\left(
\frac{\partial f[\vec{x}]}{\partial x^{\alpha}}\bigg | _{c}-
\bar{x}_{c}^{\alpha}\right)=0 \nonumber \\
\frac{\delta S}{\delta \bar{x}^{\alpha}}\bigg | _{\vec{x}_{c},
\vec{\bar{x}}_{c}} &=& -N \left( -x_{c}^{\alpha} +
\frac{1}{Q}\frac{\delta Q}{\delta \bar{x}^{\alpha}}\bigg| _{\vec
{x}_{c},\vec{\bar{x}}_{c}}\right) = 0
\label{eq59}
\end{eqnarray}
We denote the value of the action at the saddle point by $S_c$.
We have therefore the system of equations
\begin{eqnarray}
\bar{x}_{c}^{\alpha} 
&=& \frac{\partial f[\vec{x}]}{\partial x^{\alpha}}\bigg |_{c}
\label{eq60}
\end{eqnarray}
\begin{eqnarray}
x_{c}^{\alpha} = \langle \Theta^{\alpha}  \rangle, & 1\leq\alpha\leq h-1 
\label{eq61}
\end{eqnarray}
where we defined 
\begin{eqnarray}
\langle (\cdot) \rangle = 
\frac{{\rm{Tr}} 
(\cdot) e^{t[\sum_{\alpha\ne\beta=1}^{h}\mu_{\alpha\beta}\tau^{\alpha\beta}
+\sum_{\alpha=1}^{h-1}\bar{x}_{c}^{\alpha}\Theta^{\alpha}]}}
{{\rm{Tr}} e^{t[\sum_{\alpha\ne\beta=1}^{h}\mu_{\alpha\beta}\tau^{\alpha\beta}
+\sum_{\alpha=1}^{h-1}\bar{x}_{c}^{\alpha}\Theta^{\alpha}]}}
\label{eq62}
\end{eqnarray}

After this saddle-point analysis, we obtain a general expression
for the mean fitness $f_{m}$ of the population, for an arbitrary
microscopic fitness function $f(\vec{x})$,
\begin{eqnarray}
f_{m} = \lim_{N,t\rightarrow\infty}\frac{-S_{c}}{Nt}=
\max_{\{\vec{x}_{c},\vec{\bar{x}}_{c}\}}\left[f(\vec{x}_{c})-
\vec{\bar{x}}_{c}\cdot\vec{x}_{c}
-\sum_{\alpha\ne\beta=1}^{h}\mu_{\alpha\beta} 
+\lambda_{\max}\right]
\label{eq62a}
\end{eqnarray}
with $\lambda_{\max}$ defined as
\begin{eqnarray}
\lambda_{\rm{max}}=\lim_{t\rightarrow\infty}\frac{\ln Q}{t}
\label{eq62b}
\end{eqnarray}
and corresponding to the largest eigenvalue of the matrix
\begin{eqnarray}
{\mathbf{M}}(\vec{\bar{x}}_{c},\{\mu_{\alpha\beta}\})
=\sum_{\alpha\ne\beta=1}^{h}\mu_{\alpha\beta}\tau^{\alpha\beta}
+\sum_{\alpha=1}^{h-1}\bar{x}_{c}^{\alpha}\Theta^{\alpha}
\label{eq62c}
\end{eqnarray}

As shown in detail in Appendix 1, the compositions $\bar{x}^{\alpha}_{c}$
can be eliminated to reduce Eq.\ (\ref{eq62a}) to the final expression
\begin{eqnarray}
f_{m}^{(h)} &=& \max_{\{x^{1}_{c},x^{2}_{c},\ldots,x^{h-1}_{c}\}}\bigg\{
f(x^{1}_{c},x^{2}_{c},\ldots,x^{h-1}_{c}) + \sum_{\alpha\ne\beta =1}^{h-1}
 \mu_{\alpha \beta}
\left[\sqrt{x^{\alpha}_{c}x^{\beta}_{c}}-x^{\alpha}_{c}\right]\nonumber\\
&&+\sum_{\alpha=1}^{h-1}\mu_{\alpha h}\left[\sqrt{x^{\alpha}_{c}
\left(1-\sum_{\gamma=1}^{h-1}x^{\gamma}_{c}\right)} 
- x^{\alpha}_{c}\right]\nonumber\\
&&+\sum_{\beta=1}^{h-1}\mu_{h\beta}\left[\sqrt{
\left(1-\sum_{\gamma=1}^{h-1}x^{\gamma}_{c}\right)x^{\beta}_{c}}
+\sum_{\gamma = 1}^{h-1}x^{\gamma}_{c} - 1\right]\bigg\}
\label{eqGePar20gen_1}
\end{eqnarray}
Here, the compositions $0 \le x_{c}^{\alpha} \le 1$
 are defined within the simplex
$\sum_{\alpha=1}^{h-1}x_{c}^{\alpha}\leq 1$.

\subsection{The purine/pyrimidine alphabet $h$=2}

Most analytical and numerical studies of quasi-species
models have been formulated in the past \cite{Park06,Park07,Baake01,Saakian06} 
by using a
coarse-grained alphabet of nucleotides, where the
nucleotide bases are lumped into purines and pyrimidines,
and hence $h=2$. The maximum principle for this binary
alphabet can be derived from the general expression
(\ref{eqGePar20gen_1}), by assuming a symmetric mutation
rate $\mu_{12}=\mu_{21}=\mu$, and by noticing that a
single composition $x_{c}^{1}\equiv x_{c}$ 
(or normalized Hamming distance) is required in this case,
\begin{eqnarray}
f_{m}^{(2)}=\max_{\{0\leq x_{c}\leq 1\}}\bigg\{
f(x_{c}) + \mu [2\sqrt{x_{c}(1-x_{c})} - 1] 
\bigg\}
\label{eqPar_binary_1}
\end{eqnarray}

It is customary to use in this case a magnetization
coordinate \cite{Park06,Park07,Baake01,Saakian06} 
defined as $\xi_{c} = 1 - 2 x_{c}$, and hence
Eq.\ (\ref{eqPar_binary_1}) becomes
\begin{eqnarray}
f_{m}^{(2)} = \max_{\{-1\leq\xi_{c}\leq 1 \}}
\bigg\{
f(\xi_{c}) + \mu\sqrt{1-\xi_{c}^{2}} -\mu
\bigg\}
\label{eqPar_binary_2}
\end{eqnarray}

Eq.\ (\ref{eqPar_binary_2}) is a well known result, which
has been obtained in the past with different methods \cite{Saakian06,Baake01},
including a version of the Schwinger
boson method employed in the present work \cite{Park06}. It
is a special case of Eq.\ (\ref{eqGePar20gen_1}).

\subsection{The nucleic acid alphabet $h=4$}

In the nucleic acids RNA or DNA, the alphabet
is constituted by the monomers of these
polymeric chains, which are $h=4$
\begin{figure}
\centering
\includegraphics[scale=0.3]{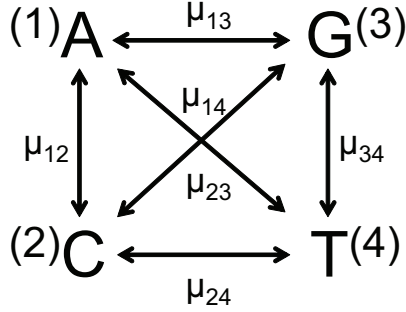}
\caption{The generalized mutation scheme}
\label{fig2}
\end{figure}
different nucleotides A, C, G, T/U. The general
mutation scheme displayed in Fig.\ \ref{fig2}
is represented by setting $h=4$ in our
general solution Eq.\ (\ref{eqGePar20gen_1}), with
3 independent compositions (or normalized Hamming distances)
$x^{1}_{c}$, $x^{2}_{c}$, $x^{3}_{c}$
\begin{eqnarray}
f_{m} &=& \max_{\{x^{1}_c,x^{2}_c,x^{3}_c\}}\bigg\{
f(x^{1}_c,x^{2}_c,x^{3}_c) + \sum_{\alpha\ne\beta=1}^3 \mu_{\alpha\beta}
\left [\sqrt{x^{\alpha}_c x^{\beta}_c }- x^{\alpha}_c \right]\nonumber\\
&&+\sum_{\alpha = 1}^{3}\mu_{\alpha 4}\left[\sqrt{x^{\alpha}_c
\left(1-\sum_{\gamma=1}^{3}x^{\gamma}_c \right)} - x^{\alpha}_c\right]
\nonumber\\
&&+\sum_{\beta=1}^{3}\mu_{4\beta}\left[\sqrt{\left(1-\sum_{\gamma=1}^{3}
x^{\gamma}_c \right)  x^{\beta}_{c}
}
+\sum_{\gamma=1}^{3}x^{\gamma}_c - 1\right]
\bigg\}
\label{eqGePar_1}
\end{eqnarray} 
Here, the compositions $x_{c}^{\alpha}$ are defined
within the simplex $x^{1}_{c}+x^{2}_{c}+x^{3}_{c}\leq 1$.

\begin{figure}[tbp]
\centering
\includegraphics[scale=0.3]{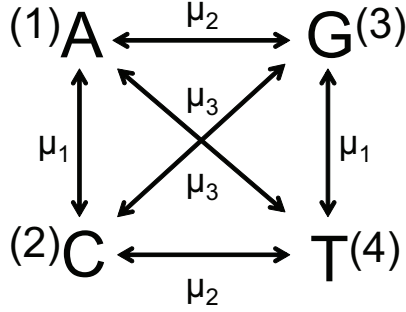}
\caption{The Kimura 3 ST mutation scheme}
\label{fig1}
\end{figure}

An interesting reduction of this
general model is provided by the Kimura 3 ST 
mutation scheme \cite{Kimura81,Hermisson01,Garske03,Garske04}, 
Fig.\ \ref{fig1}. 
The Kimura 3 ST mutation scheme considers mutations 
in three independent directions, with rates $\mu_{1},\mu_{2},\mu_{3}$
\begin{displaymath}
X \begin{array}{l}
\nearrow\raise5pt\hbox{$\mu_{1}$} \\
\rightarrow \mu_{2} \\
\searrow\lower5pt\hbox{$\mu_{3}$}
\end{array}
\end{displaymath}

Accordingly, three components of the Hamming distance between
a pair of sequences $S_{i}$ and $S_{j}$ are defined as follows
\begin{eqnarray}
d_{1}(S_{i},S_{j}) = \#_{A\leftrightarrow C}(S_{i},S_{j}) + 
\#_{G\leftrightarrow T}(S_{i},S_{j}) \nonumber
\\
d_{2}(S_{i},S_{j}) = \#_{A\leftrightarrow G}(S_{i},S_{j}) +
\#_{C\leftrightarrow T}(S_{i},S_{j})
\\
d_{3}(S_{i},S_{j}) = \#_{A\leftrightarrow T}(S_{i},S_{j}) +
\#_{C\leftrightarrow G}(S_{i},S_{j}) \nonumber
\label{eq3}
\end{eqnarray}
Here,
$\#_{X\leftrightarrow Y}(S_{i},S_{j})$ is the number of sites
at which X and Y are exchanged between sequences $S_{i}$ and $S_{j}$. The
total Hamming distance between sequences $S_{i}$ and $S_{j}$ is
given by
\begin{eqnarray}
d(S_{i},S_{j}) = d_{1}(S_{i},S_{j}) + d_{2}(S_{i},S_{j}) + 
d_{3}(S_{i},S_{j})
\label{eq4}
\end{eqnarray}

The mutation rate is therefore modeled by the function
\begin{eqnarray}
\mu_{ij} = \left\{ \begin{array}{ll}
\mu_{\alpha}, & d_{\alpha}(S_{i},S_{j})=d(S_{i},S_{j})=1\\
-N\sum_{\alpha=1}^{3}\mu_{\alpha}, & S_{i} = S_{j}\\
0, & d(S_{i},S_{j}) > 1
\end{array} \right.
\label{eq5}
\end{eqnarray}

To be consistent with existing notation in the quasispecies literature,
we define 3 independent variables, which are simply transformations of
the composition variables, and which are called `surplus' variables
in the literature
\begin{eqnarray}
u_{1}(S_{i}) = 1 - \frac{2}{N}[d_{1}(S_{i},S_{w}) + d_{3}(S_{i},S_{w})]
\nonumber \\
u_{2}(S_{i}) = 1 - \frac{2}{N}[d_{2}(S_{i},S_{w}) + d_{3}(S_{i},S_{w})]
\\ 
u_{3}(S_{i}) = 1 - \frac{2}{N}[d_{1}(S_{i},S_{w}) + d_{2}(S_{i},S_{w})]
\nonumber
\label{eq7}
\end{eqnarray}

Notice that according to this definition, the maximum value
of $u = (u_{1} + u_{2} + u_{3})/3= 1$ 
is reached when $d_{1}=d_{2}=d_{3}=0$, that is the sequence
$S_{i}$ being identical to the wild type $S_{w}$. The minimum value for 
the average base composition is 
obtained when one of the Hamming distance components, say
$d_{i}=N$, while the others are null $d_{j\neq i}=0$. Then, $d=d_i=N$ and
$u=-1/3$.    

The Kimura 3ST mutation scheme
result is obtained from the general Eq.\ (\ref{eqGePar_1})
if the following symmetries are assumed for the mutation rates
\begin{eqnarray}
\mu_{12}&=&\mu_{21}=\mu_{34}=\mu_{43}\equiv\mu_{1}\nonumber\\
\mu_{13}&=&\mu_{31}=\mu_{24}=\mu_{42}\equiv\mu_{2}\nonumber\\
\mu_{14}&=&\mu_{41}=\mu_{23}=\mu_{32}\equiv\mu_{3}
\label{eqGePar_2}
\end{eqnarray}
We follow the quasispecies literature convention
and define the 3 independent `ancestral distribution' 
coordinates $\xi^{\alpha}_{c}$ (the subscript c denoting the
saddle-point limit), after Eq.\ (\ref{eq7})
\begin{eqnarray}
\xi_{c}^{1} &=& 1 - 2(x^{1}_c + x^{3}_c)\nonumber\\
\xi_{c}^{2} &=& 1 - 2(x^{2}_c + x^{3}_c)\nonumber\\
\xi_{c}^{3} &=& 1 - 2(x^{1}_c + x^{2}_c)\nonumber\\
\label{eqGePar_3}
\end{eqnarray}
The ancestral distribution variable $\xi$ is defined as
the steady state analog of the `surplus,' but for the
time-reversed evolution process
\cite{Hermisson02}.

After some algebra, we obtain
\begin{eqnarray}
f_{m} &=& \max_{\{\xi_{c}^{1},\xi_{c}^{2},\xi_{c}^{3}\}}\bigg\{
f(\xi_{c}^{1},\xi_{c}^{2},\xi_{c}^{3}) - (\mu_{1}+\mu_{2}+\mu_{3})
\nonumber\\
&&+\frac{\mu_{1}}{2}[\sqrt{(1+\xi_{c}^{1}+\xi_{c}^{2}+\xi_{c}^{3})(1-\xi_{c}^{1}-\xi_{c}^{2}+\xi_{c}^{3})}\nonumber\\
&&+\sqrt{(1+\xi_{c}^{1}-\xi_{c}^{2}-\xi_{c}^{3})(1-\xi_{c}^{1}
+\xi_{c}^{2}-\xi_{c}^{3})}]
\nonumber\\
&&+\frac{\mu_{2}}{2}[\sqrt{(1+\xi_{c}^{1}+\xi_{c}^{2}+\xi_{c}^{3})(1+\xi_{c}^{1}-\xi_{c}^{2}-\xi_{c}^{3})}\nonumber\\
&&+\sqrt{(1-\xi_{c}^{1}+\xi_{c}^{2}-\xi_{c}^{3})(1-\xi_{c}^{1}
-\xi_{c}^{2}+\xi_{c}^{3})}]\nonumber\\
&&+\frac{\mu_{3}}{2}[\sqrt{(1+\xi_{c}^{1}+\xi_{c}^{2}+\xi_{c}^{3})(1-\xi_{c}^{1}+\xi_{c}^{2}-\xi_{c}^{3})}\nonumber\\
&&+\sqrt{(1+\xi_{c}^{1}-\xi_{c}^{2}-\xi_{c}^{3})(1-\xi_{c}^{1}
-\xi_{c}^{2}+\xi_{c}^{3})}]\bigg\}
\label{eq63}
\end{eqnarray}
From this general expression, the average composition `surplus'
$\vec{u}=(u_{1},u_{2},u_{3})$ is obtained by applying the self-consistent
condition $f(\vec{u}) = f_{m}$. This result is equivalent to that
derived by \cite{Garske04}.

\subsection{Analytic results for the symmetric mutational scheme}

For a symmetric mutational scheme, $\mu_{1}=\mu_{2}=\mu_{3}\equiv\mu$,
we specialize the general Eq.\ (\ref{eq63}) by setting
$\xi_{c}^{1}=\xi_{c}^{2}=\xi_{c}^{3}\equiv\xi_{c}$, and $u_i = u$,  
and thus obtaining an expression for the mean fitness
\begin{eqnarray}
f_{m}=\max_{-\frac{1}{3}\leq\xi_{c}\leq 1}\left\{
f(\xi_{c})-\frac{3}{2}\mu(1+\xi_{c})
+\frac{3}{2}\mu\sqrt{(1-\xi_{c})(1+3\xi_{c})}\right\}\nonumber\\
\label{eq67}
\end{eqnarray}
This result is equivalent to that derived by \cite{Garske04}.
We remark that Eq.\ (\ref{eq67}) represents an exact analytical
expression for the mean fitness of the population, for 
any arbitrary microscopic fitness $f(u)$, with the assumption
of symmetric mutation rates $\mu_{1}=\mu_{2}=\mu_{3}=\mu$.
From this exact expression, the average composition $u$ is
obtained by applying the self-consistency condition $f_{m} = f(u)$.
In the following sections, we apply Eq.\ (\ref{eq67}) to analyze
in detail some examples of microscopic fitness functions: The
sharp peak landscape, a Fujiyama landscape, a quadratic
fitness landscape, and a quartic fitness landscape.
We note that the $h=2$ case contains a symmetry in the mutation terms
about $\xi = 0$.  In the general $h>2$, this symmetry will be lost.  As
we will see, loss of this symmetry leads to a change in the order
of the error catastrophe phase transition.

\subsubsection{The sharp peak landscape}

\begin{figure}[tbp]
\centering
\includegraphics[scale=0.4]{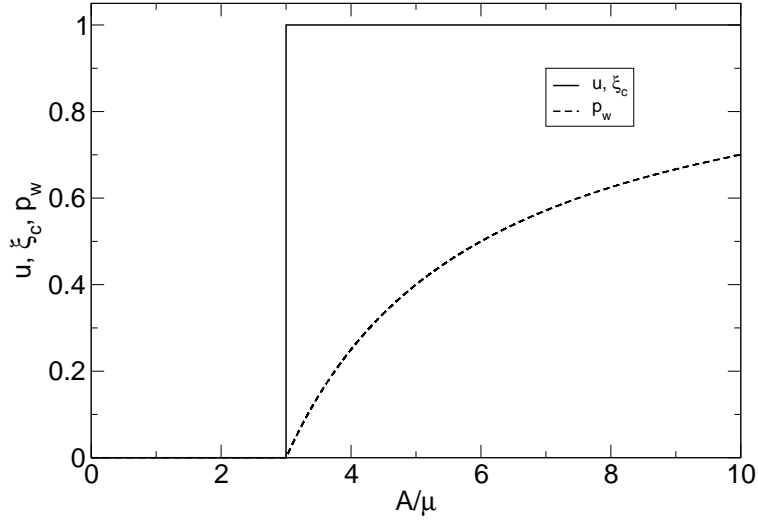}
\caption{Average composition $u$, magnetization $\xi_{c}$ and
fraction of the population at the wild-type sequence $p_{w}$,
as a function of the parameter $A/\mu$, for the sharp peak
fitness.}
\label{fig4_par_sharp}
\end{figure}

We shall first consider the sharp peak landscape, which is described
by the function
\begin{eqnarray}
f(u)=A\delta_{u,1}
\label{eq68}
\end{eqnarray}
That is, only sequences identical to the wild-type replicate with
a rate $A>0$. 
From Eq. (\ref{eq67}), we notice that this implies: $\xi_{c}=1$, 
if $A>3\mu$, or $\xi_{c}=0$ otherwise. Therefore, we obtain
for the mean replication rate
\begin{eqnarray}
f_{m}=\left\{\begin{array}{cc}A-3\mu, & A>3\mu\\
0, & A < 3\mu \end{array}\right.
\label{eq69}
\end{eqnarray}
The fraction of the population at the wild-type $p_{w}$ is 
obtained from the self-consistent condition $f_{m}=p_{w}A$,
\begin{eqnarray}
p_{w}=\left\{\begin{array}{cc}1-\frac{3 \mu}{A}, & A>3\mu\\
0, & A < 3\mu \end{array}\right.
\label{eq70}
\end{eqnarray}
There exists an error threshold in this case, which is given by the
critical value $A_{crit}=3\mu$, as shown in Eqs. (\ref{eq69}), (\ref{eq70})
and displayed in Fig.\ \ref{fig4_par_sharp}. The phase transition
is first order as a function of $A/\mu$. 

One may compare this result with the 
error threshold observed in the binary alphabet case, which 
is \cite{Park06} $A_{crit}=\mu$.
This result is intuitive, because in the 4 letters alphabet, there
exist 3  mutation channels to escape from the wild type instead of
just one as in the binary alphabet, and therefore
a stronger selection pressure is required to retain the wild-type
features.

\subsubsection{The Fujiyama fitness landscape}

The Fujiyama landscape is obtained as a linear function
of the composition
\begin{eqnarray}
f[\vec{u}]=\alpha_{1}u_{1}+\alpha_{2}u_{2}+
\alpha_{3}u_{3}
\label{eq71}
\end{eqnarray}
We will present analytical results for the symmetric case 
$\alpha_{i}\equiv\alpha$, $\mu_{i}\equiv\mu$. Thus,
$\xi_{c}^{1}=\xi_{c}^{2}=\xi_{c}^{3}=\xi_{c}$.
Substituting in Eq.\ (\ref{eq67}), we have
\begin{eqnarray}
f_{m}=\max_{-\frac{1}{3}\leq\xi_{c}\leq 1}\left\{
3\alpha\xi_{c}-\frac{3}{2}\mu(1+\xi_{c})
+\frac{3}{2}\mu\sqrt{(1-\xi_{c})(1+3\xi_{c})}\right\}
\label{eq72}
\end{eqnarray}
We look for a maximum
\begin{eqnarray}
\frac{\partial f_{m}}{\partial\xi_{c}}= 3\alpha - \frac{3}{2}\mu
+\frac{3}{2}\mu\frac{2-6\xi_{c}}{2\sqrt{1+2\xi_{c}-3\xi_{c}^{2}}}=0
\label{eq73}
\end{eqnarray}
From this equation, we obtain
\begin{eqnarray}
\xi_{c}=\frac{1}{3}\left(1+\frac{2\alpha-\mu}
{\sqrt{\alpha^{2}-\alpha\mu+\mu^{2}}} \right)
\label{eq74}
\end{eqnarray}
To obtain the average base composition $u$, 
we apply the self-consistent condition
$f_{m}=f(u)=3\alpha u$, to obtain
\begin{eqnarray}
u=\frac{1}{3}\left(1-2\frac{\mu}{\alpha}+\frac{2}{\alpha}
\sqrt{\alpha^2-\mu\alpha
+\mu^{2}} \right)
\label{eq75}
\end{eqnarray}
Clearly, no phase transition is observed in this fitness landscape,
as $0<u<1$ for $0<\alpha<\infty$. This result is in agreement with
the analysis presented in \cite{Hermisson01}, were a quantum spin
chain formulation was employed.

\subsubsection{Quadratic fitness landscape}
The quadratic fitness landscape is given by the general quadratic
form
\begin{eqnarray}
f(\vec{u})=\sum_{i=1}^{3}(\frac{\beta_{i}}{2}u_{i}^{2}+\alpha_{i}u_{i})
\label{eq76}
\end{eqnarray}

We will present the analytical solution for 
the symmetric case $\alpha_{i} \equiv \alpha$, 
$\beta_{i}\equiv\beta$, with the symmetric mutation scheme
$\mu_{i} \equiv \mu$. Under these conditions, we have $\xi_{c}^{1}=
\xi_{c}^{2} = \xi_{c}^{3} \equiv \xi_{c}$, and from Eq.\ (\ref{eq67}) 
we have for the mean fitness
\begin{eqnarray}
f_{m}=\max_{-\frac{1}{3}\leq\xi_{c}\leq 1}\left\{\frac{3}{2}\beta\xi_{c}^{2}
+3\alpha\xi_{c}-\frac{3}{2}\mu (1+\xi_{c})+\frac{3}{2}\mu
\sqrt{(1-\xi_{c})(1+3\xi_{c})} \right\}
\label{eq77}
\end{eqnarray} 
The maximum is obtained from the equation
\begin{eqnarray}
\frac{\partial f_{m}}{\partial \xi_{c}}=3\beta\xi_{c}+ 3\alpha - \frac{3}{2}\mu
+\frac{3}{2}\mu\frac{2-6\xi_{c}}{2\sqrt{1+2\xi_{c}-3\xi_{c}^{2}}}=0
\label{eq78}
\end{eqnarray}

From Eq.\ (\ref{eq78}), we obtain 
\begin{eqnarray}
\beta\xi_{c}+\alpha -\frac{\mu}{2}=\frac{\mu}{2}\frac{3\xi_{c}-1}
{\sqrt{1+2\xi_{c}-3\xi_{c}^{2}}} 
\label{eq79}
\end{eqnarray}
As shown in Appendix 2, this equation can be cast in the form of a quartic
equation, whose roots are the values of $\xi_{c}$. The average 
composition $u$
is finally obtained through the self-consistency equation
\begin{eqnarray}
f_{m} = f(u) = \frac{3}{2}\beta u^{2}+3\alpha u
\label{eq80}
\end{eqnarray}

We find that the error threshold transition 
towards a selective phase for $\alpha=0$ is defined
by $\xi_c > 0$, $u>0$, at $\beta > 3/2\mu$. The
value of $u$ is continuous at the transition, as it is straightforward
to check from Eq.\ (\ref{eq78}) $f_{m}(\xi_{c}=0)=f_{m}(\xi_{c}=2/3,\beta=3\mu/2)=0$, which implies after Eq.\ (\ref{eq80}) (for $\alpha=0$) that
$u=0$ when approaching the critical point $\beta=3\mu/2$ from both sides. 
However,
$\xi_c$ jumps from 0 [for $\beta \rightarrow (3\mu/2)^{-}$] to 2/3
[for $\beta \rightarrow (3\mu/2)^{+}$] (Appendix 2). 
To analyze the order
of the transition, we expand Eq.\ (\ref{eq77}) as a quadratic
polynomial in $\xi_{c}$ in a neighborhood of 
the critical point $\beta \simeq 3\mu/2$,
\begin{eqnarray}
f_{m}(\xi_{c}) = \left\{ \begin{array}{cc}
\frac{3}{2}(\beta - 3\mu/2)\xi_{c}^{2} & ,\beta < 3\mu/2\\
\frac{2}{3}\beta - \mu + (2\beta - 3\mu)(\xi_{c}-\frac{2}{3})
+(\frac{3}{2}\beta-3\mu)(\xi_{c}-\frac{2}{3})^2 & ,\beta > 3\mu/2
\end{array}\right.\nonumber\\
\label{eq80b}
\end{eqnarray} 
we find that the first derivative 
$df_{m}/d\beta[\beta\rightarrow (3\mu/2)^{-}]=0$, while 
$df_{m}/d\beta[\beta\rightarrow (3\mu/2)^{+}]=2/3$, and thus it
has a discontinuous
jump from $0$ to $2/3$. Therefore, the phase transition is first
order as a function of $\beta/\mu$. We notice that the order
of the phase transition, for a similar quadratic fitness
landscape, is found to be of second order for a binary
alphabet \cite{Park06}. 

When 
$0\leq\alpha/\beta\leq\frac{1}{3}\left(\sqrt{\frac{4}{3}}-1\right)$, as
shown in Appendix 2, we find a finite jump in the magnetization
from $\xi_{c,+}$ to $\xi_{c,-}$, with 
$\xi_{c,\pm} = 1/3(1\pm\sqrt{1-18\alpha/\beta - 27(\alpha/\beta)^2})$.
This result is in agreement with \cite{Hermisson01}, where
an alternative method of quantum spin chains was applied for
the derivation. A
complete analysis of the different possible cases other than this,
is presented in Appendix 2. 

\begin{figure}[tbp]
\centering
\includegraphics[scale=0.4]{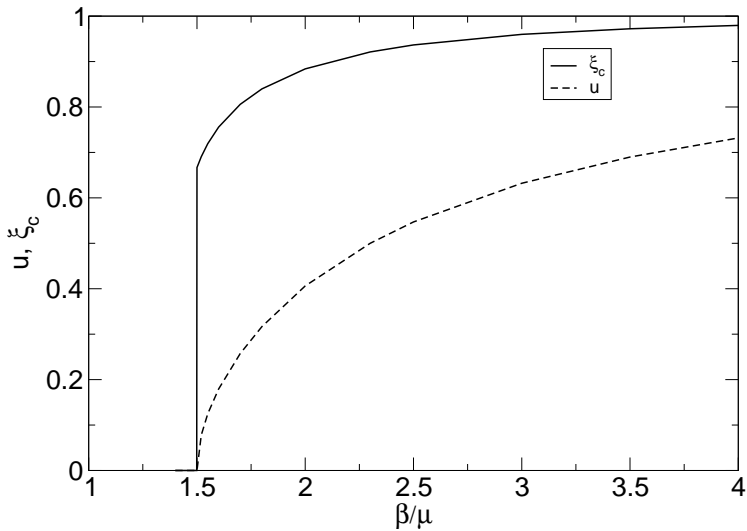}
\caption{Average composition $u$ and magnetization $\xi_{c}$
as a function of the parameter $\beta/\mu$, for the quadratic
fitness when $\alpha = 0$.}
\label{fig4_1}
\end{figure}

\subsection{Quartic fitness landscape}

As a final example, we consider a quartic fitness landscape
\begin{eqnarray}
f(\vec{u}) = \sum_{i=1}^{3}\frac{\beta_{i}}{4}u_{i}^{4}
\label{eqq1}
\end{eqnarray}
As in the previous cases, we consider the symmetric mutation
rates $\mu_{i}\equiv\mu$, $\beta_{i}\equiv\beta$, 
and hence $\xi_{c,i}\equiv\xi_{c}$.
Considering this fitness function in the general equation (\ref{eq67}), 
we have that the mean fitness is given by the analytical expression
\begin{eqnarray}
f_{m}=\max_{\{-\frac{1}{3}\leq\xi_{c}\leq 1\}}\bigg\{\frac{3}{4}\beta
\xi_{c}^{4} - \frac{3}{2}\mu(1 + \xi_{c}) 
+ \frac{3}{2}\mu\sqrt{(1-\xi_{c})(1+3\xi_{c})}\bigg\}
\label{eqq2}
\end{eqnarray}
The average composition $u$ is obtained by applying the self-consistent
condition 
\begin{eqnarray}
f(u) = \frac{3}{4}\beta u^{4} = f_{m}
\label{eqq3}
\end{eqnarray}

\begin{figure}[tbp]
\centering
\includegraphics[scale=0.4]{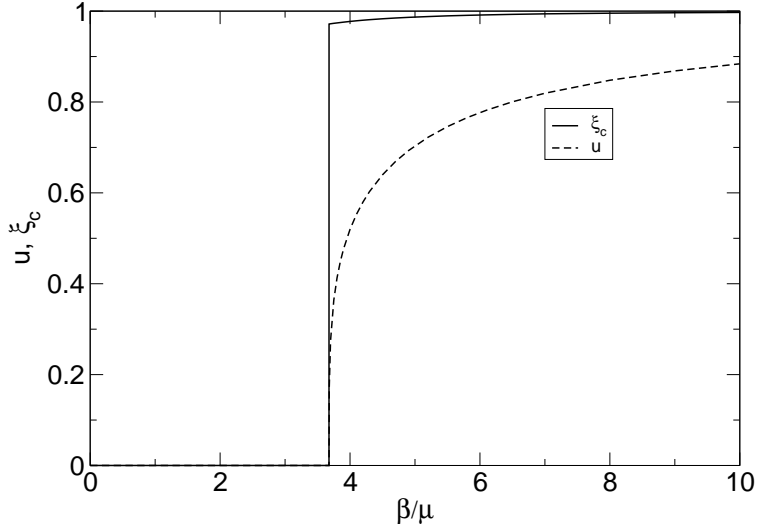}
\caption{Average composition $u$ and magnetization $\xi_{c}$
as a function of the parameter $\beta/\mu$, for the quartic
fitness landscape.}
\label{fig4_2}
\end{figure}

In Fig.\ \ref{fig4_2}, 
we present the values of $u$ and $\xi_{c}$, as obtained
from Eqs.\ (\ref{eqq2}), (\ref{eqq3}), as a function of the
parameter $\beta/\mu$. A discontinuous
jump in the bulk magnetization from $\xi_{c}=0$ to
$\xi_{c}=0.971618$ is observed at $\beta/\mu = 3.67653$. By expanding
Eq.\ (\ref{eqq2}) near the critical point, after similar
procedure as in the quadratic fitness case, we find a discontinuous
jump
in the derivative $df_{m}/d\beta$, from $0$ to $0.66841101$. Therefore,
the phase transition is first order in $\beta/\mu$. 
The average composition $u$, however, experiences a fast
but smooth transition.
This
behavior is much alike the one observed in the sharp peak
fitness landscape, Eq.\ (\ref{eq69}) and Fig.\ (\ref{fig4_par_sharp}),
except for the fact that the average composition $u$ is continuous
at the transition.
Indeed, from a purely
mathematical perspective, a fitness function following a power
law $f_{n}(u) = k u^{n}$, for $0 < u <1$, will satisfy the
limit
\begin{eqnarray}
\lim_{n\rightarrow\infty} f_{n}(u) = k\delta_{u,1}
\label{eqq4}
\end{eqnarray}
which is precisely the sharp peak landscape, Eq.\ (\ref{eq68}).
 
\subsection{Symmetric  case, general h, with
application to the amino acid alphabet $h=20$}

An alternative language to describe molecular evolution is in terms
of mutation and selection acting over the translated protein sequence,
which is drawn from an $h=20$ amino acid alphabet. For the parallel model,
the time evolution of an infinite population of protein sequences is
described by the system of differential equations (\ref{eq2}), with $h=20$.
Thus, we are lead to consider the $h=20$ case.  We first consider
the symmetric case with general $h$.

For an alphabet of arbitrary size $h$, a symmetrical mutation scheme
$\mu_{\alpha\beta} = \mu$, and a symmetrical fitness function that
leads to $x^{\alpha}_c= x_c$, we define a magnetization
coordinate $\xi_{c} = 1 - h x_c$, and Eq.\ (\ref{eqGePar20gen_1}) reduces
to
\begin{eqnarray}
f_{m} &=& \max_{\{-1/(h-1) \leq \xi_{c} \leq 1 \}}\bigg\{
f(\xi_{c}) + (h-1)\mu[\frac{2}{h}\sqrt{(1-\xi_{c})(1+(h-1) \xi_{c})} 
\nonumber\\
&&+ \frac{h-2}{h}(1-\xi_{c}) - 1]\bigg\}
\label{eqGePar20_1}
\end{eqnarray}

As an example of application of Eq.\ (\ref{eqGePar20_1}), we consider
the sharp peak fitness landscape $f(\xi_{c}) = A\delta_{\xi_{c},1}$.
Then, from Eq.\ (\ref{eqGePar20_1}) we obtain the mean
fitness
\begin{eqnarray}
f_{m} = \left\{\begin{array}{cc}A - (h-1)\mu, & A > (h-1)\mu\\
0, & A \leq (h-1)\mu\end{array}\right.
\label{eqGePar20_2}
\end{eqnarray}
We obtain the fraction of sequences in the wild type, $p_{w}$, by
applying the self-consistency condition $f_{m} = A p_{w}$, which
yields
\begin{eqnarray}
p_{w} = \left\{\begin{array}{cc}1 - \frac{(h-1)\mu}{A}, & A > (h-1)\mu\\
0, & A \le (h-1)\mu\end{array}\right.
\label{eqGePar20_4}
\end{eqnarray}
This result is intuitive, since there exists $h-1$ independent
mutation channels for the sequence to escape from
the wild type. Moreover, for a general alphabet of size $h$, 
a first order phase transition occurs at the
critical point $A_{\rm{crit}}^{h} = (h-1)\mu$.

As a second example, we consider the quadratic fitness landscape
for an alphabet of size $h$, $f(\xi_{c})=(h-1)\beta\xi_{c}^{2}/2$. 
For the quadratic fitness function we can work out the order of
the phase transition for general $h$.  We 
consider $ f(\xi) = (h-1)\beta \xi^2/2$.  There is a phase
transition at $\beta_{crit} = 2 \mu (h-1)/h$.  The
magnetization jumps  from $\xi_{c} = 0$ at $\beta \rightarrow \beta_{crit}^{-}$
to $\xi_{c} = (h-2)/(h-1)$ at $\beta \rightarrow \beta_{crit}^{+}$.
The first derivative at the critical point is:
\begin{eqnarray}
\frac{ df_m}{d\beta} = \left\{\begin{array}{cc} 
                  0, & \beta < 2 \mu (h - 1) / h \\
                  (h - 2)^2 /[2 (h - 1)], & \beta > 2 \mu (h - 1) / h
  \end{array}\right.
\label{add1}
\end{eqnarray}
Thus, the jump in the first derivative is $(h - 2)^2 / [2(h - 1)]$.
Thus, the transition is second order for $h=2$ and first order for
$h > 2$.

\section {The $h$-states Eigen model}
The Eigen model conceptually differs from the parallel or Kimura model
because it is assumed that mutations arise as a consequence of errors
in the replication process. For an alphabet of size $h$, the system of 
equations which describes 
the time evolution of the probabilities $p_{i}$, with $1 \leq i \leq h^{N}$,
is 
\begin{eqnarray}
\frac{dp_{i}}{dt}= \sum_{j=1}^{N}[B_{ij}r_{j}-\delta_{ij}D_{j}]p_{j}
-p_{i}\left[\sum_{j=1}^{N}(r_{j}-D_{j})p_{j} \right]
\label{eq81}
\end{eqnarray} 
Here, $r_{i}$ is the replication rate of sequence $S_i$, 
and the components of the matrix
\begin{eqnarray}
B_{ij}=(1-q)^{N-d_{ij}}\left(\frac{q}{h-1}\right)^{d_{ij}}
\label{eq82}
\end{eqnarray}
represent the transition rates from sequence
$S_j$ into $S_i$, where $1-q$ is the probability to copy
a nucleotide without error, 
and $q$ is the probability per site for a base substitution 
during the
replication process.

We consider a generalized version of this, by considering that
the base substitution probabilities are not necessarily identical
nor symmetric. That is, for a base substitution $\beta \rightarrow \alpha$,
we consider a probability $q_{\alpha\beta}\ne q_{\beta\alpha}$.
Correspondingly, we define $h-1$ independent
base compositions $0\le x^{\alpha}\le 1$, which can also be interpreted
as normalized Hamming distances with respect to the $h$-th reference
species, $x^{\alpha} = d^{\alpha}/N$. For this generalized
mutation scheme, the transition rate matrix components are
defined by
\begin{eqnarray}
B_{ij} = (1-q)^{N(1-\sum_{\alpha=1}^{h-1}x^{\alpha})}
\prod_{\alpha\ne\beta=1}^{h}\left(q_{\alpha\beta}\right)^{N x^{\alpha}}
\label{eq82b}
\end{eqnarray}
Here, $q = \sum_{\alpha\ne\beta=1}^{h}q_{\alpha\beta}$, and the
$x^\alpha$ are as in Sec.\ 2.1.

\subsection{The $h$-states Eigen model in operator form}
By similar arguments as in the parallel model, we formulate a
Hamiltonian operator for the Eigen model
\begin{eqnarray}
-\hat{H} &=& \prod_{j=1}^{N}\left[(1-q)\vec{\hat{a}}^{\dagger}(j)
\cdot \vec{\hat{a}}(j) + \sum_{\alpha\ne\beta=1}^{h}q_{\alpha\beta}
\vec{\hat{a}}^{\dagger}(j)\tau^{\alpha\beta}\vec{\hat{a}}(j)\right]
\nonumber \\
&&\times Nf\left[\frac{1}{N}\sum_{l=1}^{N}\vec{\hat{a}}^{\dagger}(l)
\mathbf{\Theta}\vec{\hat{a}}(l) \right]
-Nd\left[\frac{1}{N}\sum_{l=1}^{N}\vec{\hat{a}}^{\dagger}(l)
\mathbf{\Theta}\vec{\hat{a}}(l) \right]
\label{eq84}
\end{eqnarray}
Here, the matrices $\tau^{\alpha\beta}$ are defined as in the parallel
model by Eq.\ (\ref{eq17})
and in the matrix array 
${\mathbf{\Theta}}=(\Theta^{1},\Theta^{2},\ldots,\Theta^{h-1})$, the
matrices $\Theta^{\alpha}$ are defined as in Eq.\ (\ref{eq26}).
Let us define the coefficients $\mu_{\alpha\beta}=N q_{\alpha\beta}$. 
The degradation function is given by $D_i = N d(x^1, x^2, \ldots,
x^{h-1})$.
Then, we have for $q \ll 1$
\begin{eqnarray}
-N\ln(1-q)\simeq N q&=&N\sum_{\alpha\ne\beta=1}^{h}q_{\alpha\beta}
= \sum_{\alpha\ne\beta=1}^{h}\mu_{\alpha\beta} 
\label{eq85}
\end{eqnarray}

The Hamiltonian operator Eq.\ (\ref{eq84}) is expressed, to
$\mathcal{O}(1/N)$, by
\begin{eqnarray}
-\hat{H} 
&=& Ne^{-\sum_{\alpha\ne\beta=1}^{h}\mu_{\alpha\beta}}
e^{\sum_{j=1}^{N}\left[\sum_{\alpha\ne\beta=1}^{h}
\frac{\mu_{\alpha\beta}}{N}\vec{\hat{a}}^{\dagger}(j)\tau^{\alpha\beta}
\vec{\hat{a}}(j)\right]} \nonumber \\
&&\times f\left[\frac{1}{N}\sum_{l=1}^{N}\vec{\hat{a}}^{\dagger}(l)
\mathbf{\Theta}\vec{\hat{a}}(l) \right]
-Nd\left[\frac{1}{N}\sum_{l=1}^{N}\vec{\hat{a}}^{\dagger}(l)
\mathbf{\Theta}\vec{\hat{a}}(l) \right]
\label{eq86}
\end{eqnarray}

To study the equilibrium properties of the system, as in the case
of the parallel model, we calculate the partition function by
performing a Trotter factorization
\begin{eqnarray}
Z &=& {\rm{Tr}}\: e^{-\hat{H}t}\hat{P} \nonumber \\
&=&\int_{0}^{2\pi}\left[\prod_{j=1}^{N}\frac{d\lambda_{j}}{2\pi}\right]
e^{-i\lambda_{j}}\lim_{M\rightarrow\infty}
\int\left[\prod_{k=1}^{M}\prod_{j=1}^{N}
\frac{d^{h}\vec{z}_{k}^{*}(j)d^{h}\vec{z}_{k}(j)}{\pi^{h}}\right]
e^{-S[\vec{z}^{*},\vec{z}]}\nonumber\\
\label{eq87}
\end{eqnarray}
Here,
\begin{eqnarray}
e^{-S[\vec{z}^{*},\vec{z}]}=\prod_{k=1}^{M}
\langle \{\vec{z}_{k}\}|e^{-\epsilon\hat{H}}|\{\vec{z}_{k-1}\}\rangle
\label{eq88}
\end{eqnarray}
where the matrix elements in the coherent states basis 
are given by the expression
\begin{eqnarray}
\langle \{\vec{z}_{k}\}|e^{-\epsilon\hat{H}}|\{\vec{z}_{k-1}\}\rangle
&=&e^{-\frac{1}{2}\sum_{j=1}^{N}(\vec{z}^{*}_{k}(j)\cdot\vec{z}_{k}(j)
-2\vec{z}_{k}^{*}(j)\cdot\vec{z}_{k-1}(j)
+\vec{z}_{k-1}^{*}(j)\cdot\vec{z}_{k-1}(j))}
\nonumber \\
&&\times\exp\bigg(\epsilon Ne^{-\sum_{\alpha\ne\beta=1}^h\mu_{\alpha\beta}}
e^{\frac{1}{N}\sum_{j=1}^{N}\vec{z}_{k}^{*}(j)
\left(\sum_{\alpha\ne\beta=1}^{h}\mu_{\alpha\beta}\tau^{\alpha\beta}\right)
\vec{z}_{k-1}(j)}
 \nonumber \\
&&\times f\left[\frac{1}{N}\sum_{j=1}^{N}\vec{z}^{*}_{k}(j)
\mathbf{\Theta}\vec{z}_{k-1}(j) \right] 
 -\epsilon N d\left[\frac{1}{N}\sum_{j=1}^{N}\vec{z}^{*}_{k}(j)
\mathbf{\Theta}\vec{z}_{k-1}(j) \right]\bigg)
\nonumber\\
\label{eq89}
\end{eqnarray}

Let us introduce the $(h-1)$-component vector field
$\vec{x}_{k}=(x_{k}^{1},x_{k}^{2},\ldots,x_{k}^{h-1})$, and 
an integral representation
of the corresponding delta function
\begin{eqnarray}
1 &=& \int\mathcal{D}[\vec{x}]\prod_{k=1}^{M}\delta^{(h-1)}
\left[\vec{x}_{k}-\frac{1}{N}\sum_{j=1}^{N}
\vec{z}_{k}^{*}(j)\mathbf{\Theta}\vec{z}_{k-1}(j)\right]
\nonumber \\
&=& \int\left[\prod_{k=1}^{M}\prod_{\alpha=1}^{h-1}\frac{i\epsilon N 
d\bar{x}_{k}^{\alpha}
d x_{k}^{\alpha}}{2\pi}\right]e^{-\epsilon N 
\sum_{k=1}^{M}\vec{\bar{x}}_{k}
\cdot\vec{x}_{k}+\epsilon\sum_{k=1}^{M}\sum_{j=1}^{N}
\vec{z}^{*}_{k}(j)\vec{\bar{x}}_{k}\cdot\mathbf{\Theta}
\vec{z}_{k-1}(j) }
\nonumber\\
\label{eq90}
\end{eqnarray}
Similarly, let us introduce a second set of fields $\eta^{\alpha\beta}$,
\begin{eqnarray}
1&=&\int\mathcal{D}[\eta^{\alpha\beta}]\prod_{k=1}^{M}
\delta\left[\eta_{k}^{\alpha\beta}-\frac{1}{N}\sum_{j=1}^{N}\vec{z}_{k}^{*}(j)
\tau^{\alpha\beta}\vec{z}_{k-1}(j)\right] \nonumber \\
&=&\int\left[\prod_{k=1}^{M}
\frac{i\epsilon N d \bar{\eta}_{k}^{\alpha\beta}
d\eta_{k}^{\alpha\beta}}{2\pi}\right] \nonumber \\
&&\times e^{-\epsilon N\sum_{k=1}^{M}
\bar{\eta}_{k}^{\alpha\beta}\eta_{k}^{\alpha\beta}+\epsilon\sum_{k=1}^{M}
\bar{\eta}_{k}^{\alpha\beta}\sum_{j=1}^{N}\vec{z}_{k}^{*}(j)
\tau^{\alpha\beta}\vec{z}_{k-1}(j)}\nonumber\\
\label{eq91}
\end{eqnarray}
Inserting both constraints Eq.\ (\ref{eq90}) and Eq.\ (\ref{eq91}) 
in the expression for the trace Eq.\ (\ref{eq87}), we obtain
\begin{eqnarray}
Z &=& \lim_{M\rightarrow \infty}\int\mathcal{D}[\vec{x}]
\mathcal{D}[\vec{\bar{x}}]\prod_{\alpha\ne\beta}
\mathcal{D}[\eta^{\alpha\beta}]
\mathcal{D}[\bar{\eta}^{\alpha\beta}] \nonumber \\
&&\times e^{\epsilon N \sum_{k=1}^{M}(-\vec{\bar{x}}_{k}\cdot
\vec{x}_{k}-\sum_{\alpha\ne\beta=1}^{h}\bar{\eta}_{k}^{\alpha\beta}
\eta_{k}^{\alpha\beta}
+ 
e^{\sum_{\alpha\ne\beta=1}^{h}\mu_{\alpha\beta}(\eta^{\alpha\beta}-1)}
f[\vec{x}_{k}]-d[\vec{x}_{k}])} \nonumber \\
&&\times \int\mathcal{D}[\vec{z}^{*}]\mathcal{D}[\vec{z}]
\mathcal{D}[\lambda]\prod_{j=1}^{N}e^{-i\lambda_{j}}
e^{\epsilon\sum_{k,l=1}^{M}\vec{z}_{k}^{*}(j)S_{kl}\vec{z}_{l}(j)}
\bigg |_{\{\vec{z}_{0}\}=\{e^{i\lambda_{j}}\vec{z}_{M}\}}
\label{eq92}
\end{eqnarray}
After performing the Gaussian integral over the fields $\vec{z}^{*},
\vec{z}$, we obtain
\begin{eqnarray}
Z &=& \lim_{M\rightarrow \infty}\int\mathcal{D}[\vec{x}]
\mathcal{D}[\vec{\bar{x}}]
\prod_{\alpha\ne\beta}\mathcal{D}[\eta^{\alpha\beta}]
\mathcal{D}[\bar{\eta}^{\alpha\beta}] \nonumber \\
&&\times e^{\epsilon N \sum_{k=1}^{M}(-\vec{\bar{x}}_{k}\cdot
\vec{x}_{k}-\sum_{\alpha\ne\beta=1}^h \bar{\eta}_{k}^{\alpha\beta}
\eta_{k}^{\alpha\beta}
+ e^{-\sum_{\alpha\ne\beta=1}^h \mu_{\alpha\beta}(1-\eta^{\alpha\beta}_{k})}
f[\vec{x}_{k}]-d[\vec{x}_{k}])
}
 \nonumber \\
&&\times \mathcal{D}[\lambda]\prod_{j=1}^{N}e^{-i\lambda_{j}}
[\det S(j)]^{-1}
\label{eq93}
\end{eqnarray}
The matrix $S(j)$ has the structure
\begin{eqnarray}
S(j) = \left(\begin{array}{ccccc}
I & 0 & 0 & \ldots & -e^{-i\lambda_{j}}A_{1}(j) \\
-A_{2}(j) & I & 0 & \ldots & 0 \\
0 & -A_{3}(j) & I & \ldots & 0 \\
  &  &  & \ddots & \\
0 & & \ldots & -A_{M}(j) & I \end{array} \right)
\label{eq94}
\end{eqnarray}
where 
$A_{k}(j) = I + \epsilon[
\sum_{\alpha\ne\beta=1}^{h}\mu_{\alpha\beta}\bar{\eta}^{\alpha\beta}_{k}
+ \vec{\bar{x}}_{k}\cdot\mathbf{\Theta}]$. We
obtain
\begin{eqnarray}
\det S(j) &=& \det \left[ I - e^{i\lambda_{j}}\prod_{k=1}^{M}A_{k}(j)\right]
\nonumber \\
&=&\det\left[ I - e^{i\lambda_{j}}\hat{T}e^{\epsilon\sum_{k=1}^{M}
[\sum_{\alpha\ne\beta=1}^{h}\bar{\eta}^{\alpha\beta}_{k}\tau^{\alpha\beta}
+\vec{\bar{x}}_{k}\cdot
\mathbf{\Theta}]}\right] \nonumber \\
&=& e^{{\rm{Tr}} \ln \left[ I - e^{i\lambda_{j}}
\hat{T}e^{\epsilon\sum_{k=1}^{M}
[\sum_{\alpha\ne\beta=1}^{h}\bar{\eta}^{\alpha\beta}_{k}\tau^{\alpha\beta}
+\vec{\bar{x}}_{k}\cdot
\mathbf{\Theta}]}\right]}
\label{eq95}
\end{eqnarray} 
Substituting this last expression into the functional integral 
Eq.\ (\ref{eq93}),
and then performing the integrals over the $\lambda$ fields, we obtain
\begin{eqnarray}
Z&=&\lim_{M\rightarrow\infty}\int\mathcal{D}[\vec{\bar{x}}]
\mathcal{D}[\vec{x}]\prod_{\alpha\ne\beta=1}^h
\mathcal{D}[\bar{\eta}^{\alpha\beta}]
\mathcal{D}[\eta^{\alpha\beta}] \nonumber \\
&&\times e^{\epsilon N\sum_{k=1}^{M}(
-\vec{\bar{x}}_{k}\cdot\vec{x}_{k}-
\sum_{\alpha\ne\beta=1}^h \bar{\eta}_{k}^{\alpha\beta}\eta_{k}^{\alpha\beta}+
e^{\sum_{\alpha\ne\beta=1}^h \mu_{\alpha\beta}(\eta^{\alpha\beta}-1)}
f[\vec{x}_{k}]-d[\vec{x}_{k}])}\prod_{j=1}^{N}Q
\nonumber\\
\label{eq96}
\end{eqnarray}
Here,
\begin{eqnarray}
Q &=&\lim_{M\rightarrow\infty}{\rm{Tr}}\,\hat{T}
\prod_{k=1}^{M}[I+\epsilon(
\sum_{\alpha\ne\beta=1}^h \bar{\eta}_{k}^{\alpha\beta}\tau^{\alpha\beta}
+\vec{\bar{x}}_{k}\cdot\mathbf{\Theta})] \nonumber \\
&=&\lim_{M\rightarrow\infty} 
{\rm{Tr}}\,\hat{T}e^{\epsilon\sum_{k=1}^{M}[
\sum_{\alpha\ne\beta=1}^{h}\bar{\eta}_{k}^{\alpha\beta}\tau^{\alpha\beta}
+\vec{\bar{x}}_{k}\cdot\mathbf{\Theta}]} \nonumber \\
&=&{\rm{Tr}}\,\hat{T}e^{\int_{0}^{t}dt'[
\sum_{\alpha\ne\beta=1}^{h}\bar{\eta}^{\alpha\beta}(t')\tau^{\alpha\beta}
+\vec{\bar{x}}(t')\cdot\mathbf{\Theta}]}
\label{eq97}
\end{eqnarray}
After taking the limit $M\rightarrow\infty$, we obtain
\begin{eqnarray}
Z=\int\mathcal{D}[\vec{\bar{x}}]
\mathcal{D}[\vec{x}]\prod_{\alpha\ne\beta}
\mathcal{D}[\bar{\eta}^{\alpha\beta}]
\mathcal{D}[\eta^{\alpha\beta}]
e^{-S[\vec{\bar{x}},\vec{x},
\{\bar{\eta}^{\alpha\beta}\},\{\eta^{\alpha\beta}\}]}
\label{eq98}
\end{eqnarray}
Here,
\begin{eqnarray}
S[\vec{\bar{x}},\vec{x},
\{\bar{\eta}^{\alpha\beta}\},\{\eta^{\alpha\beta}\}]&=&
-N\int_{0}^{t}dt'\{-\vec{\bar{x}}(t')
\cdot\vec{x}(t')-\sum_{\alpha\ne\beta=1}^{h}
\bar{\eta}^{\alpha\beta}(t')\eta^{\alpha\beta}(t')
\nonumber \\
&&+e^{\sum_{\alpha \ne \beta=1}^{h}\mu_{\alpha\beta}
(\eta^{\alpha  \beta}(t')-1)}
f[\vec{x}(t')]-d[\vec{x}(t')]\}-N\ln Q
\nonumber\\
\label{eq99}
\end{eqnarray}
where
\begin{eqnarray}
Q = 
{\rm{Tr}}\:
\hat{T}e^{\int_{0}^{t}dt'[
\sum_{\alpha\ne\beta=1}^{h}\bar{\eta}^{\alpha\beta}(t')\tau^{\alpha\beta}
+ \sum_{\alpha=1}^{h-1}\bar{x}^{\alpha}(t')\Theta^{\alpha}]} 
\label{eq101}
\end{eqnarray}
With this last simplification, the effective action becomes
\begin{eqnarray}
S[\vec{\bar{x}},\vec{x},
\{\bar{\eta}^{\alpha\beta}\},\{\eta^{\alpha\beta}\}]&=&
-N\int_{0}^{t}dt'(-\vec{\bar{x}}(t')
\cdot\vec{x}(t')- \sum_{\alpha\ne\beta=1}^{h}\bar{\eta}^{\alpha\beta}(t')
\eta^{\alpha\beta}(t')
\nonumber \\
&&+e^{\sum_{\alpha\ne\beta=1}^{h}\mu_{\alpha\beta}(\eta^{\alpha\beta}-1)}
f[\vec{x}(t')]-d[\vec{x}(t')])-N\ln Q
\nonumber\\
\label{eq102}
\end{eqnarray}

\subsection{The large N limit of the $h$-state Eigen model is a saddle point}
By assuming that the sequence length N is very large, 
$N\rightarrow\infty$, we can evaluate
the functional integral Eq. (\ref{eq98}) 
by a saddle point method. Considering the
action defined in Eq. (\ref{eq102}), we have
\begin{eqnarray}
\frac{\delta S}{\delta x^{\alpha}}\bigg |_{\vec{\bar{x}}_{c},
\vec{x}_{c},\{\bar{\eta}_{c}^{\alpha\beta}\},\{\eta_{c}^{\alpha\beta}\}}
&=&N\left(\bar{x}_{c}^{\alpha}-
e^{\sum_{\gamma\ne\rho=1}^h \mu_{\gamma\rho}(\eta^{\gamma\rho}_{c}-1)}
\frac{\partial f[\vec{x}]}{\partial x^{\alpha}}\bigg |_{c}+
\frac{\partial d[\vec{x}_{c}]}{\partial x^{\alpha}}\bigg |_{c} 
\right)=0 \nonumber \\
\frac{\delta S}{\delta \eta^{\alpha\beta}}\bigg |_{\vec{\bar{x}}_{c},
\vec{x}_{c},\{\bar{\eta}_{c}^{\alpha\beta}\},\{{\eta}_{c}^{\alpha \beta}\}}
&=&N\left(\bar{\eta}_{c}^{\alpha\beta}-\mu_{\alpha\beta}
e^{\sum_{\gamma\ne\rho=1}^h \mu_{\gamma\rho}(\eta^{\gamma\rho}_{c}-1)}
f[\vec{x}_{c}] \right)=0 
\nonumber \\
\frac{\delta S}{\delta \bar{x}^{\alpha}}\bigg |_{\vec{\bar{x}}_{c},
\vec{x}_{c},\{\bar{\eta}^{\alpha\beta}_{c}\},
\{\eta_{c}^{\alpha\beta}\}}
&=&N\left(x_{c}^{\alpha}-\frac{1}{Q}\frac{\delta Q}{\delta\bar{x}^{\alpha}}
\bigg |_{c} 
\right)=0
\nonumber \\
\frac{\delta S}{\delta \bar{\eta}^{\alpha\beta}}
\bigg |_{\vec{\bar{x}}_{c},
\vec{x}_{c},\{\bar{\eta}_{c}^{\alpha\beta}\},
\{\eta_{c}^{\alpha\beta}\}}
&=&
N\left(\eta_{c}^{\alpha \beta}-\frac{1}{Q}\frac{\delta Q}{\delta\bar{\eta}^{\alpha \beta}}
\bigg |_{c} 
\right)=0
\label{eq103}
\end{eqnarray}
We have therefore the system of equations
\begin{eqnarray}
\bar{x}_{c}^{\alpha}=e^{\sum_{\gamma\ne\rho=1}^h \mu_{\gamma\rho}(
\eta_{c}^{\gamma\rho}-1)}
\frac{\partial f[\vec{x}]}{\partial x^{\alpha}}\bigg |_{c}-
\frac{\partial d[\vec{x}_{c}]}{\partial x^{\alpha}}\bigg |_{c} 
\label{eq104}
\end{eqnarray}
\begin{eqnarray}
\bar{\eta}_{c}^{\alpha\beta}=\mu_{\alpha\beta}
e^{\sum_{\gamma\ne\rho=1}^h \mu_{\gamma\rho}(\eta_{c}^{\gamma\rho}-1)}
f[\vec{x}_{c}]
\label{eq105}
\end{eqnarray}
\begin{eqnarray}
x_{c}^{\alpha} = \langle\Theta^{\alpha}\rangle 
\label{eq106}
\end{eqnarray}
\begin{eqnarray}
\eta_{c}^{\alpha\beta} = \langle\tau^{\alpha\beta}\rangle 
\label{eq107}
\end{eqnarray}
where we defined 
\begin{eqnarray}
\langle (\cdot) \rangle = 
\frac{{\rm{Tr}}\:  (\cdot)
e^{t[\sum_{\alpha\ne\beta=1}^{h}\bar{\eta}^{\alpha\beta}_{c}\tau^{\alpha\beta}
+ \sum_{\alpha=1}^{h-1}\bar{x}^{\alpha}_{c}\Theta^{\alpha}]}}
{{\rm{Tr}}\: e^{t[\sum_{\alpha\ne\beta=1}^{h}\bar{\eta}_{c}^{\alpha\beta}
\tau^{\alpha\beta}
+ \sum_{\alpha=1}^{h-1}\bar{x}^{\alpha}_{c}\Theta^{\alpha}]}}
\nonumber\\
\label{eq108}
\end{eqnarray}

After the saddle-point analysis, we obtain an exact analytical
expression for the mean fitness $f_{m}$ of the population, in the limit
of very large sequences $N\rightarrow\infty$, for an
arbitrary microscopic fitness function $f(\vec{x})$ and
degradation rate $d(\vec{x})$
\begin{eqnarray}
f_{m} = \lim_{N,t\rightarrow\infty}\frac{-S_{c}}{Nt}&=&
\max_{\{\vec{x}_{c}\vec{\bar{x}}_{c},\{\eta_{c}^{\alpha\beta}\},
\{\bar{\eta}_{c}^{\alpha\beta}\}\}}
[e^{\sum_{\alpha\ne\beta=1}^{h}\mu_{\alpha\beta}(\eta^{\alpha\beta}_{c}-1)}
f(\vec{x}_{c})-d(\vec{x}_{c})\nonumber\\
&&-\vec{\bar{x}}_{c}
\cdot\vec{x}_{c}-\sum_{\alpha\ne\beta=1}^{h}\eta^{\alpha\beta}_{c}
\bar{\eta}^{\alpha\beta}_{c}
+\lambda_{\max}]
\nonumber\\
\label{eq108b}
\end{eqnarray}

As shown in Appendix 3, the $\eta^{\alpha\beta}$ can be eliminated
in terms of the compositions, to obtain the final expression
\begin{eqnarray}
f_{m} &=& \max_{\{x^{1}_{c},x^{2}_{c},\ldots,x^{h-1}_{c}\}}\bigg\{
f(x^{1}_{c},x^{2}_{c},\ldots,x^{h-1}_{c}) e^{\sum_{\alpha\ne\beta=1}^{h-1}
\mu_{\alpha\beta}[\sqrt{x^{\alpha}_{c}x^{\beta}_{c}} - x^{\alpha}_{c}]}
\nonumber\\
&&\times
e^{\sum_{\alpha=1}^{h-1}\mu_{\alpha h}[\sqrt{x^{\alpha}_{c}\left(
1-\sum_{\gamma=1}^{h-1}x^{\gamma}_{c}\right)}-x^{\alpha}_{c}]}\nonumber\\
&&\times e^{\sum_{\beta=1}^{h-1}\mu_{h\beta}
[\sqrt{\left(1-\sum_{\gamma=1}^{h-1}x_{c}^{\gamma}\right)x^{\beta}_{c}}
+\sum_{\gamma = 1}^{h-1}x^{\gamma}_{c} - 1]} 
- d(x^{1}_{c},x^{2}_{c},\ldots,x^{h-1}_{c})
\bigg\}
\nonumber\\
\label{eqGeEig20_1}
\end{eqnarray}
Here, the compositions $0 \le x_{c}^{\alpha} \le 1$ 
are defined within the simplex
$\sum_{\alpha=1}^{h-1}x_{c}^{\alpha}\leq 1$.
We note that the mutation terms in Eq.\  (\ref{eqGeEig20_1}) are
the exponential of the mutation terms in Eq.\ (\ref{eqGePar20gen_1}),
which is a result of the mutation terms in the Eigen Hamiltonian, 
Eq.\ (\ref{eq86}),
being the exponential of those in the parallel Hamiltonian,
Eq.\ (\ref{eq30}).

\subsection{The purine/pyrimidine alphabet $h$=2}

As an application of our general solution Eq.\ (\ref{eqGeEig20_1}),
we first consider the purine/pyrimidine alphabet with $h=2$. The
maximum principle for this binary alphabet can be derived
from Eq.\ (\ref{eqGeEig20_1}) by assuming a symmetric mutation
rate $\mu_{12}=\mu_{21}\equiv \mu$, and by noticing that a
single composition (or normalized Hamming distance) $x_{c}^{1}\equiv x_{c}$
is required in this case,
\begin{eqnarray}
f_{m}^{(2)}=\max_{0\leq x_{c} \leq 1}\left\{
e^{\mu[2\sqrt{x_{c}(1-x_{c})}-1]} - d(x_{c})
\right\}
\label{eqGeEig_2}
\end{eqnarray}

It is customary to use in this case a magnetization coordinate, defined
as $\xi_{c}=1 - 2 x_{c}$, and hence Eq.\ (\ref{eqGeEig_2}) becomes
\begin{eqnarray}
f_{m}^{(2)} = \max_{\{-1\leq\xi_{c}\leq 1\}}\left\{
e^{\mu[\sqrt{1-\xi_{c}^{2}}-1]}f(\xi_{c}) - d(\xi_{c})\right\}
\label{eqGeEig_3}
\end{eqnarray}

Eq.\ (\ref{eqGeEig_3}) is a well known result \cite{Park06,Saakian06},
and  a special case of our general result Eq.\ (\ref{eqGeEig20_1}).

\subsection{The nucleic acid alphabet $h=4$}
For a general, non-symmetric mutation scheme as in Fig. 1,
by considering $h=4$ in our general result Eq.\ (\ref{eqGeEig20_1})
we obtain
\begin{eqnarray}
f_{m} &=& \max_{\{x^{1}_{c},x^{2}_{c},x^{3}_{c}\}}\bigg\{
f(x^{1}_{c},x^{2}_{c},x^{3}_{c})
e^{\sum_{\alpha\ne\beta=1}^{ 3}\mu_{\alpha\beta}
[\sqrt{x^{\alpha}_{c}x^{\beta}_{c}}-x^{\alpha}_{c}]}
e^{\sum_{\alpha=1}^{3}\mu_{\alpha 4}[\sqrt{x^{\alpha}_{c}
\left(1- \sum_{\gamma=1}^{3}x^{\gamma}_{c}\right)}-x^{\alpha}_{c}]}
\nonumber\\
&&\times e^{\sum_{\beta=1}^{3}\mu_{4\beta}[
\sqrt{\left(1-\sum_{\gamma=1}^{3}x_{c}^{\gamma}\right)x^{\beta}_{c}}
+ \sum_{\gamma=1}^{3}x^{\gamma}_{c} - 1]}
- d(x^{1}_{c},x^{2}_{c},x^{3}_{c})
\bigg\}
\label{eqGeEig_4}
\end{eqnarray}
Here, the compositions $x^{\alpha}_{c}$ are defined within
the simplex $x^{1}_{c}+x^{2}_{c}+x^{3}_{c}\leq 1$.

An interesting reduction of this general model is provided by the
Kimura 3ST mutation scheme, introduced in section 2.6, and
represented in Fig. 2. We obtain the solution for the Kimura 3ST
mutation scheme by assuming the following symmetries in the
mutation coefficients
\begin{eqnarray}
\mu_{12} &=& \mu_{21} = \mu_{34} = \mu_{43} \equiv \mu_{1}\nonumber\\
\mu_{13} &=& \mu_{31} = \mu_{24} = \mu_{42} \equiv \mu_{2}\nonumber\\
\mu_{14} &=& \mu_{41} = \mu_{23} = \mu_{32} \equiv \mu_{3}
\label{eqGeEig_5}
\end{eqnarray}
along with the magnetization coordinates $\xi_{c}^{\alpha}$
defined in agreement with Eq.\ (\ref{eqGeEig_5}),
\begin{eqnarray}
\xi_{c}^{1} &=& 1 - 2(x^{1}_c + x^{3}_c)\nonumber\\
\xi_{c}^{2} &=& 1 - 2(x^{2}_c + x^{3}_c)\nonumber\\
\xi_{c}^{3} &=& 1 - 2(x^{1}_c + x^{2}_c)
\label{eqGeEig_6}
\end{eqnarray}
With these assumptions, after some algebra, Eq.\ (\ref{eqGeEig20_1})
reduces to the expression for the Kimura 3ST scheme,
\begin{eqnarray}
f_{m} &=& \max_{\{\xi^{1}_{c},\xi^{2}_{c},\xi^{3}_{c}\}}\bigg\{
e^{\mu_{1}(\frac{1}{2}\sqrt{(1+\xi_{c}^{1}+\xi_{c}^{2}+\xi_{c}^{3})
(1-\xi_{c}^{1}-\xi_{c}^{2}+\xi_{c}^{3})}+\frac{1}{2}
\sqrt{(1+\xi_{c}^{1}-\xi_{c}^{2}-\xi_{c}^{3})(1-\xi_{c}^{1}+\xi_{c}^{2}
-\xi_{c}^{3})}-1)}\nonumber\\
&&\times e^{\mu_{2}(\frac{1}{2}\sqrt{(1+\xi_{c}^{1}+\xi_{c}^{2}+\xi_{c}^{3})
(1+\xi_{c}^{1}-\xi_{c}^{2}-\xi_{c}^{3})})
+\frac{1}{2}\sqrt{(1-\xi_{c}^{1}+\xi_{c}^{2}-\xi_{c}^{3})(1-\xi_{c}^{1}
-\xi_{c}^{2}+\xi_{c}^{3})}-1)}\nonumber\\
&&\times e^{\mu_{3}(\frac{1}{2}\sqrt{(1+\xi_{c}^{1}+\xi_{c}^{2}+\xi_{c}^{3})
(1-\xi_{c}^{1}+\xi_{c}^{2}-\xi_{c}^{3})}
+\frac{1}{2}\sqrt{(1+\xi_{c}^{1}-\xi_{c}^{2}-\xi_{c}^{3})
(1-\xi_{c}^{1}-\xi_{c}^{2}+\xi_{c}^{3})}-1)}
\nonumber\\
&&\times f(\xi_{c}^{1},\xi_{c}^{2},\xi^{3}_{c}) 
- d(\xi^{1}_{c},\xi^{2}_{c},\xi^{3}_{c})
\bigg\}
\label{eqGeEig_7}
\end{eqnarray}

\subsection{Analytical results for the symmetric mutation scheme}
If the mutation rates are identical $\mu_{1}=\mu_{2}=\mu_{3}=\mu$,
then we have the symmetric case 
$\xi_{c}^{1}=\xi_{c}^{2}=\xi_{c}^{3}\equiv\xi_{c}$, and
after Eq.\ (\ref{eqGeEig_7}) we obtain 
\begin{eqnarray}
f_{m}=\max_{-\frac{1}{3}\leq\xi_{c}\leq 1}
\left\{
e^{\frac{3}{2}\mu
[-1-\xi_{c}+\sqrt{(1+3\xi_{c})(1-\xi_{c})}]}f[\xi_{c}]-d[\xi_{c}]\right\}
\label{eq118}
\end{eqnarray}

\subsubsection{The sharp peak fitness landscape}

\begin{figure}[tbp]
\centering
\includegraphics[scale=0.4]{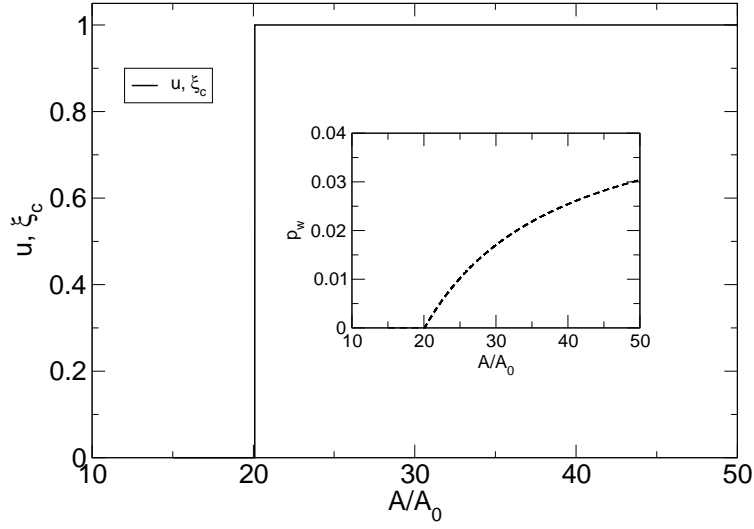}
\caption{The average composition $u$ and magnetization $\xi_{c}$
are represented as a function of the parameter $A/A_{0}$, for the
sharp peak landscape. The mutation rate was set to $\mu=1.0$. Also
shown (inset) is the fraction of the population located at the peak,
$p_{w}$.}
\label{fig4_sharp_eig}
\end{figure}

Let us first consider the sharp peak landscape 
$f(u)=(A-A_{0})\delta_{u,1} + A_{0}$, with $A>A_0$. 
That is, the replication rate
is $f(u=1)=A$ for sequences identical to the wild type, and
$f(u\neq 1)=A_{0}$, for all other sequences. 
With zero degradation rate, 
$d=0$, we notice that this result 
implies: $\xi_c = 1$ if $A>A_0 e^{3\mu}$, or
$\xi_c = 0$ otherwise. Therefore, we obtain the mean replication rate
\begin{eqnarray}
f_{m}=\left\{\begin{array}{cc}e^{-3\mu}A,&\:A>A_0 e^{3\mu}\\
A_0,&\:A<A_0 e^{3\mu}\end{array}\right.
\label{eq119}
\end{eqnarray}
The system experiences a phase transition which is first order in $A$.
The steady-state probability for the wild-type is obtained from the
self-consistent condition: $f_{m}=A p_{w}+ A_{0}(1-p_{w})$,
\begin{eqnarray}
p_{w}=\left\{\begin{array}{cc}\frac{e^{-3\mu}A-A_{0}}{A-A_{0}},&\:
A>A_0 e^{3\mu}\\
0,&\:A<A_0 e^{3\mu}\end{array}\right.
\label{eq120}
\end{eqnarray}
Notice that the error threshold is reached at the critical value
$A_{crit}=A_0 e^{3\mu}$, as follows from Eqs.\ (\ref{eq119}), (\ref{eq120})
and as displayed in Fig.\ \ref{fig4_sharp_eig}. We notice that
this result differs from the analytical value obtained for
the binary alphabet \cite{Park06}, $A_{crit}^{(2)} = A_{0}e^{\mu}$.
The additional factor of three which is explicit in the exponent
is clearly a consequence of the existence of three mutation
channels into which evolving sequences can escape 
from the wild-type. This effect, which is purely entropic- and not
fitness-like, is an explicit consequence of the larger alphabet
size. 

\subsubsection{The Fujiyama fitness landscape}
We will consider the Fujiyama fitness landscape,
which is a linear function of the composition 
\begin{eqnarray}
f(\vec{u})=\sum_{i=1}^{3}(\alpha_{i}u_{i})+\alpha_{0}
\label{eq121}
\end{eqnarray}
For the symmetric case, 
$\alpha_{i} = \alpha$, $\mu_{i} = \mu$.
Therefore, we have $\xi_{c}^{1} = \xi_{c}^{2} = \xi_{c}^{3} \equiv \xi_{c}$.
The mean fitness, in the absence of 
degradation, from Eq.\ (\ref{eq118}) becomes
\begin{eqnarray}
f_{m}=\max_{-\frac{1}{3}\leq\xi_{c}\leq 1}\left\{(3\alpha\xi_{c}+\alpha_{0})
e^{\frac{3}{2}\mu[-1-\xi_{c}+\sqrt{(1+3\xi_{c})(1-\xi_{c})}]}\right\}
\label{eq122}
\end{eqnarray}
By maximizing with respect to $\xi_{c}$, 
$\frac{\partial f_{m}}{\partial \xi_{c}}=0$, we obtain the nonlinear
equation
\begin{eqnarray}
\alpha -\frac{\mu}{2}\alpha_{0}-\frac{3}{2}\mu\alpha\xi_{c}=
\frac{\mu}{2}\frac{(3\alpha\xi_{c}+\alpha_{0})(3\xi_{c}-1)}
{\sqrt{(1+3\xi_{c})(1-\xi_{c})}}
\label{eq123}
\end{eqnarray} 
No error threshold is observed for this fitness landscape, except for the
trivial limit $\alpha \rightarrow 0$, $\alpha_{0}\ge 0$.
The average surplus $u$ is obtained by the 
self-consistent equation
\begin{eqnarray}
f_{m}=3\alpha u + \alpha_{0}
\label{eq124}
\end{eqnarray}

\subsection{The quadratic fitness landscape}

Next we consider the quadratic fitness landscape
\begin{eqnarray}
f(\vec{u})=\sum_{i=1}^{N}\left(\frac{\beta_{i}}{2}u_{i}^{2}+\alpha_{i}u_{i}
\right) + 1
\label{eq125}
\end{eqnarray}

For the symmetric case, $\beta_{i}=\beta$, $\alpha_{i}=\alpha$, 
$\mu_{i}=\mu$, we have $\xi_{c}^{i}=\xi_{c}$ and 
$\bar{\xi}_{c}^{i}=\bar{\xi}_{c}$. Thus, the mean fitness, for a null
degradation rate, after Eq.\ (\ref{eq118}) is
\begin{eqnarray}
f_{m}=\max_{-\frac{1}{3}\leq\xi_{c}\leq 1}\left\{\left(\frac{3}{2}\beta
\xi_{c}^{2}+3\alpha\xi_{c}+1 \right)
e^{\frac{3}{2}\mu[-1-\xi_{c}+\sqrt{(1+3\xi_{c})(1-\xi_{c})}]} \right\}
\label{eq126}
\end{eqnarray}
We maximize with respect to $\xi_{c}$, 
$\frac{\partial f_{m}}{\partial\xi_{c}}=0$, to obtain
\begin{eqnarray}
\beta\xi_{c}+\alpha=\frac{\mu}{2}\left(\frac{3}{2}\beta\xi_{c}^{2}
+3\alpha\xi_{c}+1\right)\left[1+\frac{3\xi_{c}-1}
{\sqrt{(1+3\xi_{c})(1-\xi_{c})}} \right]
\label{eq127}
\end{eqnarray}
The average base composition $u$ 
is obtained from the self-consistent condition
\begin{eqnarray}
f_m = f(u) = \frac{3}{2}\beta u^{2} + 3\alpha u + 1
\label{eq128}
\end{eqnarray}

\begin{figure}[tbp]
\centering
\includegraphics[scale=0.4]{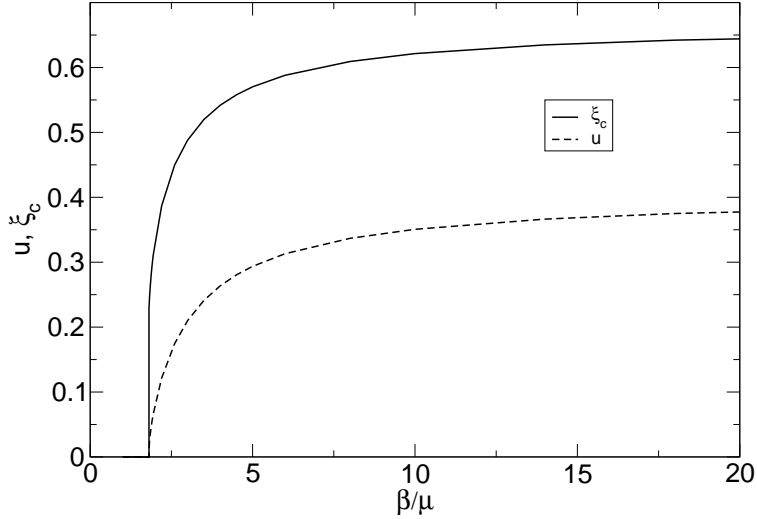}
\caption{The average composition $u$ and magnetization $\xi_{c}$
are represented as a function of the parameter $\beta/\mu$ for the
quadratic fitness,when $\alpha = 0$}
\label{fig4_3}
\end{figure}

The selected phase, $\xi_{c}>0$, $u>0$, occurs for $\beta>1.8096\mu$
when $\alpha=0$. The value of $u$ is continuous at the transition, as it can
checked from Eq.\ (\ref{eq126}) that 
$f_{m}(\xi_{c}=0)=f_{m}(\xi_{c}=0.2289,\beta=1.8096,\mu=1)=1$, which
implies after Eq.\ (\ref{eq128}) (for $\alpha=0$) 
that $u=0$ when approaching the
critical point $\beta=1.8066\mu$ from both sides. However,
$\xi_{c}$ jumps from 0 (for $\beta\rightarrow 1.8066\mu^{-}$) 
to $0.2289$ (for $\beta\rightarrow 1.8066\mu^{+}$). 
By expanding Eq.\ (\ref{eq126}) 
near the critical point, after a similar procedure as in Eq.\ (\ref{eq80b})
for the parallel model,
we find a discontinuous jump in $df_{m}/d\beta$ from
$0$ to $0.06883$. Therefore, the phase transition is of first order
in $\beta$.  A graphical representation
is displayed in Fig.\ \ref{fig4_3}.

\subsection{The quartic fitness landscape}

As a final example, we consider the quartic fitness landscape,
\begin{eqnarray}
f(\vec{u}) = \sum_{i=1}^{3}\frac{\beta_{i}}{4}u_{i}^{4} + 1
\label{eqqe1}
\end{eqnarray}
We further consider the symmetric case
$\mu_{i}\equiv\mu$, $\beta_{i}\equiv\beta$, and hence
$\xi_{c}^{i}\equiv\xi_{c}$. From the general expression Eq. (\ref{eq118}),
we obtain an analytical expression for the mean fitness
\begin{eqnarray}
f_{m} = \max_{\{-\frac{1}{3}\leq\xi_{c}\leq 1\}}\bigg\{
\left(\frac{3}{4}\beta\xi_{c}^{4}+1\right)e^{ \frac{3}{2}\mu[-1-\xi_{c}+
\sqrt{(1+3\xi_{c})(1-\xi_{c})}]}
\bigg\}
\label{eqqe2}
\end{eqnarray}
The average composition of the population $u$ is obtained from the
self-consistent condition
\begin{eqnarray}
f(u) = \frac{3}{4}\beta u^{4} + 1 = f_{m}
\label{eqqe3}
\end{eqnarray}

\begin{figure}[tbp]
\centering
\includegraphics[scale=0.4]{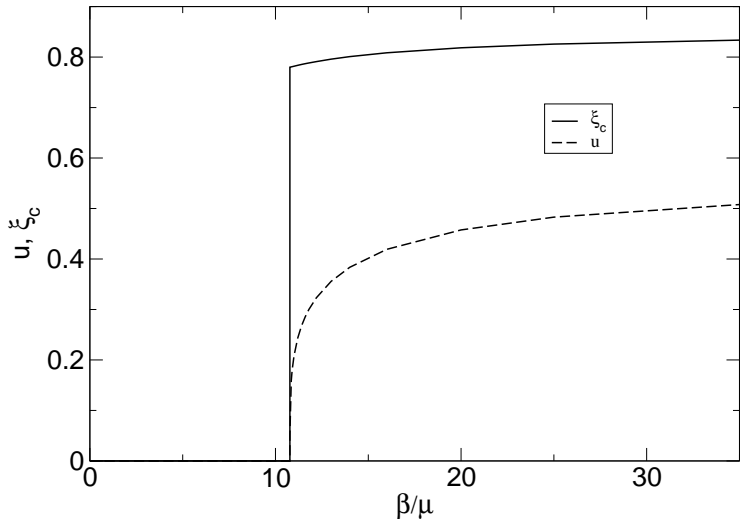}
\caption{The average composition $u$ and magnetization $\xi_{c}$
are represented as a function of the parameter $\beta/\mu$ for the
quartic fitness landscape}
\label{fig4_4}
\end{figure}

In Fig.\ \ref{fig4_4}, 
we present the values of $u$ and $\xi_{c}$, as obtained
from Eqs.\ (\ref{eqqe2}), (\ref{eqqe3}), as a function of the
parameter $\beta/\mu$. We notice that a discontinuous
jump in the bulk magnetization from $\xi_{c}=0$ to
$\xi_{c}=0.779856$ is observed at $\beta/\mu = 10.776165$. By
expanding Eq. (\ref{eqqe2}) near the critical point, we find a 
discontinuous jump in $df_{m}/d\beta$, from $0$ to $0.066213$. Therefore,
the phase transition is of first order in $\beta$. The
average composition shows a fast but
continuous transition. This
behavior is much like the one observed in the sharp peak
fitness landscape, Eq.\ (\ref{eq119}), and in the corresponding
example for the parallel model. 

\subsection{Symmetric  case, general h, with
application to the amino acid alphabet $h=20$}

We consider the case of the amino acid alphabet, which is
derived from our general solution Eq.\ (\ref{eqGeEig20_1})
by setting $h=20$. In particular, when a symmetric mutation
scheme is assumed $\mu_{\alpha\beta}= \mu$, for all $ \alpha, \beta$,
and $x^{\alpha}_{c} = x_{c}$.  We first consider the
symmetric case for general $h$.

For an alphabet of size $h$, we define a magnetization coordinate
$\xi_{c} = 1 - h x_{c}$, and obtain that Eq.\ (\ref{eqGeEig20_1})
reduces to
\begin{eqnarray}
f_{m} = \max_{\{-1/(h-1)\leq\xi_{c}\leq 1\}}
\bigg\{
f(\xi_{c}) e^{(h-1)\mu[\frac{2}{h}\sqrt{(1-\xi_{c})(1+(h-1)\xi_{c})}
+\frac{h-2}{h}(1-\xi_{c})-1]} - d(\xi_{c})
\bigg\}
\nonumber\\
\label{eqGeEig20_6}
\end{eqnarray}

As an example of application of Eq.\ (\ref{eqGeEig20_6}), we consider
the sharp peak fitness landscape 
$f(\xi_{c})=(A-A_{0})\delta_{\xi_{c},1}+A_{0}$, and zero degradation function
$d(\xi_{c})=0$. 
Then, from Eq.\ (\ref{eqGeEig20_6}) we obtain the mean fitness
\begin{eqnarray}
f_{m} = \left\{\begin{array}{cc} e^{-(h-1)\mu}A, & A > A_{0}e^{(h-1)\mu}\\
A_{0}, & A \leq A_{0}e^{(h-1)\mu}\end{array}\right.
\label{eqGeEig20_7}
\end{eqnarray}

The system experiences a first order phase transition at 
$A_{\rm{crit}}=A_{0}e^{(h-1)\mu}$. The steady-state probability for the
wild-type is obtained from the self-consistency condition: $f_{m}=A p_{w}
+ A_{0}(1 - p_{w})$,
\begin{eqnarray}
p_{w} = \left\{\begin{array}{cc}\frac{A e^{-(h-1)\mu}-A_{0}}{A-A_{0}},
& A  > A_{0}e^{(h-1)\mu}\\0, & A \leq A_{0}e^{(h-1)\mu}
\end{array}\right.
\label{eqGeEig20_9}
\end{eqnarray}
The factor $(h-1)\mu$ in the exponential is intuitive, since there exists
$h-1$ independent mutation channels into which evolving sequences can
escape from the wild-type. 

As a second example, we consider the quadratic fitness landscape
$f(\xi_{c}) = (h-1)\beta\xi_{c}^{2}/2 + 1$ for an
alphabet of size $h$. We consider the case of the amino acid alphabet,
with $h=20$. We set $\mu = 1$.
By a similar analysis as in the parallel model case,
we find a phase transition at the critical point $\beta = 7.483\mu$.
The magnetization parameter $\xi_{c}$ has a finite jump from 
$\xi_{c} = 0$ (for $\beta \rightarrow 7.483\mu^{-}$) to
$\xi_{c} = 0.0827$ (for $\beta \rightarrow 7.483\mu^{+}$). The mean
fitness is continuous at the transition, since $f_{m}(\beta\rightarrow
7.483\mu^{+}) = f_{m}(\beta\rightarrow 7.483\mu^{-}) = 1$, which
implies that the observable $u=0$ at the critical point. We observe
that the first derivative has a finite jump at the critical point, from 
$df_{m}/d\beta(\beta\rightarrow 7.483\mu^{-}) = 0$, to
$df_{m}/d\beta(\beta\rightarrow 7.483\mu^{+}) = 0.0437$, and therefore
the phase transition is of first order.
A similar analysis shows that the transition is first order
for $h=3$ as well.
As with the parallel model, we find that the transition for the quadratic
fitness function is second order for $h=2$ and  first
order for $h > 2$.

\section{Conclusion}

Using the quantum spin chain approach, the 2-state, purine/pyrimidine 
assumption for quasi-species theory has been lifted
to arbitrary alphabet sizes $h$.  
We have here expressed the 
general result for the fitness of the evolved population 
as a maximization principle.
We have derived the solution for a general fitness function 
using the Schwinger
spin coherent states approach.  We have presented analytic results 
for the sharp peak, 
as well as linear, quadratic, and quartic fitness functions.   For the
Kimura 3 ST mutation scheme, we have presented an explicit solution
for a general fitness function, expressed as a 
maximization principle.  

We have also derived the general solution to the Eigen model of mutation and
selection for arbitrary alphabet size and for a general mutation scheme.  
We have presented analytic results for the sharp peak, linear, 
quadratic, and quartic fitness functions.  

These results bring quasi-species theory closer to the evolutionary
dynamics that occurs at the genetic level.

\section*{Acknowledgments}
This work was supported by DARPA under the FunBio program,
by the AFOSR under the FAThM program, and by the
Korea Research Foundation.

\section*{Appendix 1}

In what follows, we will need to apply Euler's theorem and three properties
of the maximum eigenvalue $\lambda_{\rm{max}}$ of the matrix ${\bf{M}}$
defined by Eq.\ (\ref{eq62c}).

\subsection*{Euler's theorem for homogeneous functions}

\subsubsection*{Definition 1: Homogeneous function of degree k}
A function $f(x_{1},\ldots,x_{n})$ of $n$ variables is homogeneous
of degree $k$ if, for all $\alpha>0$,
\begin{eqnarray}
f(\alpha x_{1},\alpha x_{2},\ldots,\alpha x_{n}) =
\alpha^{k} f(x_{1},x_{2},\ldots,x_{n})
\label{eqA2_1}
\end{eqnarray}

\subsubsection*{Euler's theorem}

A differentiable function $f(x_{1},\ldots,x_{n})$ of $n$ variables is
homogeneous of degree $k$ if and only if
\begin{eqnarray}
\sum_{i=1}^{n}x_{i}\frac{\partial}{\partial x_{i}}f(x_{1},\ldots,x_{n})
=k f(x_{1},\ldots,x_{n})
\label{eqA2_2}
\end{eqnarray}

The theorem and its proof is presented in most textbooks of
mathematical analysis \cite{Fritz}.

\subsection*{Property I}

The maximum eigenvalue $\lambda_{\rm{max}}$ of the matrix ${\bf{M}}$ defined by
Eq.\ (\ref{eq62c}) is a homogeneous function of degree $k=1$ in the vector
$(\bar{x}^{1}_{c},\bar{x}_{c}^{2},\ldots,\bar{x}_{c}^{h-1}
,\{\mu_{\alpha\beta}\})$.

The proof of this proposition follows directly from Definition 1. Notice that
after Eq.\ (\ref{eq62c}),
\begin{eqnarray}
{\bf{M}}(\{\bar{x}_{c}^{\alpha}\},\{\mu_{\alpha\beta}\}) 
= \sum_{\alpha\ne\beta=1}^{h}\mu_{\alpha\beta}\tau^{\alpha\beta}
+\sum_{\alpha=1}^{h-1}\bar{x}^{\alpha}_{c}\Theta^{\alpha}
\nonumber\\
\label{eqA2_3}
\end{eqnarray}
Since ${\bf{M}}$ is a linear function of the vector 
$( \bar{x}^{1}_{c}, \ldots, \bar{x}^{h-1}_{c}, {\mu_{\alpha
 \beta}})$
\begin{eqnarray}
{\bf{M}}(\alpha\bar{x}_{c}^{1},\alpha\bar{x}_{c}^{2},\ldots,
\alpha\bar{x}_{c}^{h-1}
,\alpha\mu_{12},\alpha\mu_{21},\ldots)= \alpha {\bf{M}}(\bar{x}_{c}^{1},
\bar{x}_{c}^{2},\ldots,\bar{x}_{c}^{h-1},\mu_{12},\mu_{21},\ldots)
\nonumber\\
\label{eqA2_4}
\end{eqnarray}

Therefore, the maximum eigenvalue, as obtained from the long-time limit
of the trace,
\begin{eqnarray}
&&\lambda_{\rm{max}}(\alpha\bar{x}_{c}^{1},\alpha\bar{x}_{c}^{2},
\ldots,\alpha\bar{x}_{c}^{h-1},\alpha\mu_{12},\alpha\mu_{21},\ldots)\nonumber\\
&=&\lim_{t\rightarrow\infty}\frac{1}{t}\ln\bigg[{\rm{Tr}}\,e^{t{\bf{M}}
(\alpha\bar{x}_{c}^{1},\alpha\bar{x}_{c}^{2},\ldots,\alpha\bar{x}_{c}^{h-1},
\alpha\mu_{12},
\alpha\mu_{21},\ldots)}\bigg]\nonumber\\
&=&\lim_{t\rightarrow\infty}\frac{1}{t}\ln\bigg[{\rm{Tr}}\,e^{t\alpha{\bf{M}}(\bar{x}_{c}^{1},\bar{x}_{c}^{2},\ldots,\bar{x}_{c}^{h-1},\mu_{12},
\mu_{21},\ldots)}\bigg]\nonumber\\
&=&\alpha\lambda_{\rm{max}}(\bar{x}_{c}^{1},\bar{x}_{c}^{2},\ldots,
\bar{x}_{c}^{h-1},\mu_{12},\mu_{21},\ldots)
\label{eqA2_5}
\end{eqnarray}
is also a homogeneous function of degree $k=1$, after Definition 1.

\subsection*{Property II}
$\lambda_{\rm{max}}$ satisfies the identity
\begin{eqnarray}
\lambda_{\rm{max}} &=& 
\sum_{\alpha=1}^{h-1}\bar{x}^{\alpha}_{c}\frac{\partial \lambda_{\rm{max}}}
{\partial \bar{x}^{\alpha}_{c}}
+\sum_{\alpha\ne\beta=1}^{h}\mu_{\alpha\beta}\frac{\partial \lambda_{\rm{max}}}
{\partial\mu_{\alpha\beta}}
\label{eqA2_6}
\end{eqnarray}

The proof follows directly by application of Euler's theorem, for a
homogeneous function of degree $k=1$.

Notice that, after the saddle-point Eqs.\ (\ref{eq59})--(\ref{eq62}),
we have the following identities
\begin{eqnarray}
x_{c}^{\alpha}=\langle\Theta^{\alpha}\rangle &=&
\lim_{t\rightarrow\infty} \frac{{\rm{Tr}}\bigg[\Theta^{\alpha}e^{t{\bf{M}}} \bigg]}{{\rm{Tr}}\,e^{t{\bf{M}}}}\nonumber\\
&=&\lim_{t\rightarrow\infty}\frac{\partial}{\partial\bar{x}_{c}^{\alpha}}
\frac{\ln\bigg[{\rm{Tr}}\,e^{{t\bf{M}}}\bigg]}{t}=
\frac{\partial}{\partial\bar{x}_{c}^{\alpha}}\lambda_{{\rm{max}}}
\label{eqA2_7}
\end{eqnarray}
\begin{eqnarray}
\langle\tau^{\alpha\beta}\rangle = \frac{\partial}{\partial\mu_{\alpha\beta}}
\lambda_{{\rm{max}}}
\label{eqA2_12}
\end{eqnarray}

Substituting Eqs.\ (\ref{eqA2_7})--(\ref{eqA2_12}) 
into Eq.\ (\ref{eqA2_6}), we obtain
the identity
\begin{eqnarray}
\lambda_{{\rm{\max}}}
- \sum_{\alpha=1}^{h-1}\bar{x}_{c}^{\alpha}x_{c}^{\alpha} 
= \sum_{\alpha\ne\beta=1}^{h}\mu_{\alpha\beta}\langle\tau^{\alpha\beta}\rangle
\label{eqA2_13}
\end{eqnarray}

After Eq.\ (\ref{eqA2_13}), Eq.\ (\ref{eq62a}) becomes
\begin{eqnarray}
f_{m} = \max_{\{\bar{x}_{c}^{\alpha}\},\{\mu_{\alpha\beta}\}}
\bigg[f(\{x_{c}^{\alpha}\})
-\sum_{\alpha\ne\beta=1}^h \mu_{\alpha\beta}
(1 - \langle\tau^{\alpha\beta}\rangle)\bigg]
\nonumber\\
\label{eqA2_14}
\end{eqnarray}

\subsection*{Property III}

The 'average' $\langle {{\bf{A}}}\rangle$ of an arbitrary
$h\times h$ matrix ${{\bf{A}}}$, satisfies the identity
\begin{eqnarray}
\langle {{\bf{A}}}\rangle = \lim_{t\rightarrow\infty}
\frac{{\rm{Tr}}\bigg[{\bf{A}}\,e^{t{\bf{M}}}\bigg]}{{\rm{Tr}}\,e^{t{\bf{M}}}}
=\vec{y}_{\rm{max}}^{T}{\bf{A}}\vec{y}_{\rm{max}}
\label{eqA2_15}
\end{eqnarray}
with $\vec{y}_{\rm{max}}$ the eigenvector corresponding to the
maximum eigenvalue $\lambda_{\rm{max}}$ of the matrix ${\bf{M}}$ in
Eq.\ (\ref{eqA2_3}).

The proof follows by considering the unitary $h\times h$ matrix
${\bf{P}}=[\vec{y}_{1},\vec{y}_{2},\vec{y}_{3},\ldots,\vec{y}_{h}]$,
${\bf{P}}^{-1} = {\bf{P}}^{T}$
whose
columns are formed by the $h$ orthogonal eigenvectors $\vec{y}_{\alpha}$
of ${\bf{M}}$ which satisfy
\begin{eqnarray}{\bf{M}}\vec{y}_{\alpha} &=& \lambda_{\alpha}
\vec{y}_{\alpha} \,\,\, \, \,
1\leq \alpha \leq h\nonumber\\
\vec{y}_{\alpha}\cdot\vec{y}_{\beta} &=& \delta_{\alpha\beta}
\label{eqA2_16}
\end{eqnarray}
From elementary linear algebra, the
matrix ${\bf{P}}$ induces a similarity transformation which
diagonalizes ${\bf{M}}$, that is ${\bf{P}}{\bf{M}}{\bf{P}}^{-1}
={\rm{diag}}(\lambda_{1},\lambda_{2},\dots,\lambda_{h})
\equiv {\bf{D}}$.
Hence, it also diagonalizes the exponential of ${\bf{M}}$,
\begin{eqnarray}{\bf{P}}e^{t{\bf{M}}}{\bf{P}}^{-1} &=& {\bf{P}}({\bf{I}} + t{\bf{M}}
+ \frac{t^{2}}{2!}{\bf{M M}} + \ldots){\bf{P}}^{-1}\nonumber\\
&=& {\bf{I}} + t{\bf{P M}}{\bf{P}}^{-1} + \frac{t^{2}}{2!}{\bf{P M}}{\bf{P}}^{-1}{\bf{P M}}{\bf{P}}^{-1} + \ldots\nonumber\\
&=&{\bf{I}} + t {\bf{D}} + \frac{t^{2}}{2!}{\bf{D}}^{2} +\ldots\nonumber\\
&=& e^{t{\bf{D}}}
\label{eqA2_17}
\end{eqnarray}

The 'average' of an arbitrary $h \times h$ matrix ${\bf{A}}$, defined
by Eq.\ (\ref{eqA2_15}), is calculated as
\begin{eqnarray}
\langle{\bf{A}}\rangle = \lim_{t\rightarrow\infty}\frac{{\rm{Tr}}
\bigg[{\bf{A}}\,e^{t{\bf{M}}}\bigg]}{{\rm{Tr}}\,e^{t{\bf{M}}}}&=&
\lim_{t\rightarrow\infty}\frac{{\rm{Tr}}\bigg[{\bf{P A}}{\bf{P}}^{-1}{\bf{P}}e^{t{\bf{M}}}{\bf{P}}^{-1}\bigg]}{{\rm{Tr}}\bigg[{\bf{P}}
e^{t{\bf{M}}}{\bf{P}}^{-1}\bigg]}\nonumber\\
&=&\lim_{t\rightarrow\infty}\frac{{\rm{Tr}}\bigg[{\bf{P A}}{\bf{P}}^{-1}
e^{t{\bf{D}}} \bigg]}{{\rm{Tr}}\,e^{t{\bf{D}}}}\nonumber\\
&=&\lim_{t\rightarrow\infty}\frac{\sum_{i=1}^{h}\vec{y}_{i}^{T}{\bf{A}}
\vec{y}_{i}e^{t\lambda_{i}}}{\sum_{j=1}^{h}e^{t\lambda_{j}}}=\vec{y}_{{\rm{max}}}^{T}{\bf{A}}\vec{y}_{{\rm{max}}}
\label{eqA2_18}
\end{eqnarray}
which proves the Property III.

We can express the eigenvector
$\vec{y}_{\rm{max}}=(y^{1},y^{2},\ldots,y^{h})^{T}$
in terms of the fields $x_{c}^{\alpha}$, by
combining the result in Property III, with the saddle-point equations
Eq.\ (\ref{eqA2_7}),(\ref{eqA2_12}), as follows

\begin{eqnarray}
x_{c}^{\alpha}&=&\langle\Theta^{\alpha}\rangle = 
\vec{y}_{{\rm{max}}}^{T} \Theta^{\alpha} \vec{y}_{{\rm{max}}}\nonumber\\
&=&(y^{\alpha})^2
\label{eqA2_19}
\end{eqnarray}
These equations are inverted to obtain
\begin{eqnarray}
y^{\alpha} = \left\{\begin{array}{cc}\sqrt{x^{\alpha}_{c}}, &
1\leq\alpha\leq h-1\\\sqrt{1 - \sum_{\gamma=1}^{h-1}x^{\gamma}_{c}}, &
\alpha = h\end{array}\right.
\label{eqA2_20}
\end{eqnarray}

Equipped with this result, we can now calculate the 'averages' 
$\langle\tau^{\alpha\beta}\rangle$
in Eq.\ (\ref{eqA2_14}),
\begin{eqnarray}
\langle \tau^{\alpha\beta} \rangle &=& \vec{y}_{\rm{max}}^{T}
\tau^{\alpha\beta}\vec{y}_{\rm{max}}\nonumber\\
&=& y^{\alpha}y^{\beta} + \sum_{\gamma\ne\alpha}(y^{\gamma})^{2}
\label{eqA2_24}
\end{eqnarray}
Substituting Eq.\ (\ref{eqA2_20}) into Eq.\ (\ref{eqA2_24}), we obtain
the result
\begin{eqnarray}
\langle\tau^{\alpha\beta}\rangle = \left\{\begin{array}{cc}
\sqrt{x^{\alpha}_{c}x^{\beta}_{c}} 
+ 1 - x^{\alpha}_{c}, & \alpha\ne\beta\ne h\\
\sqrt{x^{\alpha}_{c}\left(1 - \sum_{\gamma=1}^{h-1}x^{\gamma}_{c}\right)}
+ 1 - x^{\alpha}_{c}, & \beta = h,\, \alpha\ne h\\
\sqrt{\left(1-\sum_{\gamma=1}^{h-1}x^{\gamma}_{c}\right)x^{\beta}_{c}}
+ \sum_{\gamma=1}^{h-1}x^{\gamma}_{c}, & \alpha = h,\,\beta\ne h
\end{array}\right.
\label{eqA2_25}
\end{eqnarray}

Substituting Eq.\ (\ref{eqA2_25}) into Eq.\ (\ref{eqA2_14}), we obtain
the final solution for the mean fitness of the parallel model
in an alphabet of size $h$, with an arbitrary mutation scheme,
\begin{eqnarray}
f_{m}^{(h)} &=& \max_{\{x^{1}_{c},x^{2}_{c},\ldots,x^{h-1}_{c}\}}\bigg\{
f(x^{1}_{c},x^{2}_{c},\ldots,x^{h-1}_{c}) + \sum_{\alpha\ne\beta=1}^{ h-1}
 \mu_{\alpha \beta}
\left[\sqrt{x^{\alpha}_{c}x^{\beta}_{c}}-x^{\alpha}_{c}\right]\nonumber\\
&&+\sum_{\alpha=1}^{h-1}\mu_{\alpha h}\left[\sqrt{x^{\alpha}_{c}
\left(1-\sum_{\gamma=1}^{h-1}x^{\gamma}_{c}\right)} 
- x^{\alpha}_{c}\right]\nonumber\\
&&+\sum_{\beta=1}^{h-1}\mu_{h\beta}\left[\sqrt{
\left(1-\sum_{\gamma=1}^{h-1}x^{\gamma}_{c}\right)x^{\beta}_{c}}
+\sum_{\gamma = 1}^{h-1}x^{\gamma}_{c} - 1\right]\bigg\}
\label{eqA2Par_1}
\end{eqnarray}

\section*{Appendix 2}
By performing elementary algebraic manipulations Eq. (\ref{eq79})
\begin{eqnarray}
3\beta\xi_{c} + 3\alpha -\frac{3}{2}\mu 
+ \frac{3}{2}\frac{2-6\xi_{c}}{2\sqrt{1+2\xi_{c}-3\xi_{c}^{2}}}=0
\nonumber
\end{eqnarray} 
can be cast into the standard form of a quartic equation
\begin{eqnarray}
A\xi_{c}^{4}+B\xi_{c}^{3}+C\xi_{c}^{2}+D\xi_{c}+E=0
\label{eqA8}
\end{eqnarray}
where, by defining $\tilde{\mu}\equiv \mu/\beta$ and 
$\tilde{\alpha} \equiv \alpha/\beta$, the coefficients correspond to
\begin{eqnarray}
A &=& 3 \nonumber \\
B &=& 6\tilde{\alpha}-3\tilde{\mu}-2\\
C &=& 3\tilde{\mu}^{2} - 3\tilde{\alpha}\tilde{\mu}+3\tilde{\alpha}^{2}
+2\tilde{\mu}-4\tilde{\alpha}-1 \nonumber \\
D &=& -2\tilde{\alpha}-2\tilde{\alpha}^{2}+\tilde{\mu}
+2\tilde{\alpha}\tilde{\mu}-2\tilde{\mu}^{2} \nonumber \\
E &=& -\tilde{\alpha}^{2}+\tilde{\alpha}\tilde{\mu}
\label{eqA9}
\end{eqnarray}
We remark that this quartic equation introduces additional,
unphysical solutions to the original Eq.\ (\ref{eq79}). However,
discarding these unphysical solutions whenever appropriate,
the quartic Eq.\ (\ref{eqA8}) allows us to obtain explicit
analytical expressions for $\xi_{c}$ in the entire region
of parameters. 
Following Ferrari's method \cite{Abramowitz}, we define the parameters
\begin{eqnarray}
a_1 &=& -\frac{3B^{2}}{8A^{2}}+\frac{C}{A} \nonumber \\
&=&-\frac{1}{2}-\frac{\tilde{\alpha}}{3}-\frac{\tilde{\alpha}^{2}}{2}+
\frac{\tilde{\mu}}{6}+\frac{\tilde{\alpha}\tilde{\mu}}{2}+
\frac{5\tilde{\mu}^{2}}{8}
\label{eqA10}
\end{eqnarray}
\begin{eqnarray}
a_2 &=& \frac{B^{3}}{8A^{3}}-\frac{BC}{2A^{2}}+\frac{D}{A} \nonumber \\
&=&-\frac{4}{27}-\frac{4\tilde{\alpha}}{9}+\frac{2\tilde{\mu}}{9}-
\frac{\tilde{\mu}^{2}}{4}-\frac{3\tilde{\alpha}\tilde{\mu}^{2}}{4}
+\frac{3\tilde{\mu}^{3}}{8}
\label{eqA11}
\end{eqnarray}
\begin{eqnarray}
a_3 &=& -\frac{3B^{4}}{256 A^{4}}+\frac{C B^{2}}{16A^{3}}-
\frac{B D}{4 A^{2}}+\frac{E}{A} \nonumber \\ 
&=& -\frac{5}{432}-\frac{7\tilde{\alpha}}{108}
-\frac{5\tilde{\alpha}^{2}}{72}+\frac{\tilde{\alpha}^{3}}{12}
+\frac{\tilde{\alpha}^{4}}{16}+\frac{7\tilde{\mu}}{216}+
\frac{5\tilde{\alpha}\tilde{\mu}}{72}
-\frac{\tilde{\alpha}^{2}\tilde{\mu}}{8}
-\frac{\tilde{\alpha}^{3}\tilde{\mu}}{8}+\frac{\tilde{\mu}^{2}}{288}
\nonumber \\
&&+\frac{3\tilde{\alpha}\tilde{\mu}^{2}}{16}
+\frac{9\tilde{\alpha}^{2}\tilde{\mu}^{2}}{32}
-\frac{7\tilde{\mu}^{3}}{96}-\frac{7\tilde{\alpha}\tilde{\mu}^{3}}{32}+
\frac{13\tilde{\mu}^{4}}{256}
\label{eqA12}
\end{eqnarray}
and solve the depressed quartic equation in the auxiliary variable
$z=\xi_c+B/4A$,
\begin{eqnarray}
z^4 + a_1 z^2 + a_2 z + a_3 = 0
\label{eqA13}
\end{eqnarray}

We analyze the different cases in the parameter space that
defines the possible solutions of this equation.

Case 1: $a_2 = 0$. This situation arises at the critical value
\begin{eqnarray}
\tilde{\mu}_{c}^{(1)} = \frac{2}{3} + 2\tilde{\alpha}
\label{eqA14}
\end{eqnarray}
We obtain four possible roots, according to the general formula
\begin{eqnarray}
\xi_{c}&=&-\frac{B}{4A}\pm \sqrt{\frac{-a_1\pm\sqrt{a_1^{2}-4 a_3}}{2}}
\nonumber \\
&=&\frac{1}{6}\left(2\pm\sqrt{2}\sqrt{1-9\tilde{\alpha}(2+3\tilde{\alpha}) 
\pm |1-9\tilde{\alpha}(2+3\alpha)|} \right)
\label{eqA15}
\end{eqnarray}
Depending on the sign of the term in the square root, we have the following
solutions

i)  If $1-18\tilde{\alpha}-27\tilde{\alpha}^{2} > 0$. This situation
occurs when 
$-\frac{1}{3}
\leq \tilde{\alpha} \leq\frac{1}{3}
\left(\sqrt{\frac{4}{3}}-1\right)$, and the solution is 

\begin{eqnarray}
\xi_{c,\pm} 
= \frac{1}{3}(1 \pm \sqrt{1-18\tilde{\alpha}-27\tilde{\alpha}^{2}}),
&\: \xi_{c}= \frac{1}{3} 
\label{eqA16}
\end{eqnarray}

ii) If $1-18\tilde{\alpha}-27\tilde{\alpha}^{2} \leq 0$. This situation
occurs when 
$\tilde{\alpha} > \frac{1}{3}\left(\sqrt{\frac{4}{3}}-1\right)$.
\begin{eqnarray}
\xi_{c} = \frac{1}{3}
\label{eqA17}
\end{eqnarray}
We shall consider $\tilde{\alpha}\ge 0$
in the region of physically meaningful parameters. When $\tilde{\alpha}=0$, 
a non-selective phase is obtained, from Eq. (\ref{eqA16}), if 
$\beta < \frac{3}{2}\mu$. At $\beta = \frac{3}{2}\mu$, 
for $\alpha=0$, a finite
'jump' in the value of $\xi_{c}$ from 0 to 2/3 defines a phase
transition, where the value of u varies continuously from
0 to a positive value. 

When 
$0\leq
\tilde{\alpha}\leq\frac{1}{3}\left(\sqrt{\frac{4}{3}}-1\right)$,
a finite jump in the bulk magnetization from $\xi_{c,-}$ to
$\xi_{c,+}$ is observed. This result
is in agreement with \cite{Hermisson01}.

Case 2: $a_3 = 0$, $a_2 \ne 0$. This situation occurs at the critical
values
\begin{eqnarray}
\tilde{\mu}_{c}^{(2)} = \frac{2}{39}\left(1 + 3\tilde{\alpha}+
2\sqrt{49-18\tilde{\alpha}-27\tilde{\alpha}^2} \right), & 0 \leq 
\tilde{\alpha} \leq 1.054444
\label{eqA18}
\end{eqnarray}
\begin{eqnarray}
\tilde{\mu}_{c}^{(3)} = \frac{2}{39}\left(1 + 3\tilde{\alpha}-
2\sqrt{49-18\tilde{\alpha}-27\tilde{\alpha}^2} \right), & 1 \leq 
\tilde{\alpha} \leq 1.054444 
\label{eqA18b}
\end{eqnarray}
In this case, the quartic equation in $z$ factorizes,
\begin{eqnarray}
z(z^3+a_1 z + a_2) = 0
\label{eqA19}
\end{eqnarray}

There is a solution $z=0$ for Eq. (\ref{eqA19}). This is however not
a solution of Eq. (\ref{eq79}), but an artifact of introducing
the algebraic transformation into the fourth order
polynomial Eq. (\ref{eqA8}). 

The solutions corresponding to the remaining cubic
equation in Eq. (\ref{eqA19}) are analyzed as follows. Let us define
the parameters,
\begin{eqnarray}
s_1 = \left[-\frac{a_2}{2}+\left(\frac{a_1^3}{27}+\frac{a_2^2}{4}
\right)^{1/2}\right]^{1/3}\nonumber\\
s_2 = \left[-\frac{a_2}{2}-\left(\frac{a_1^3}{27}+\frac{a_2^2}{4}
\right)^{1/2}\right]^{1/3}
\label{eqA21}
\end{eqnarray}

Then, we have the following cases,

Case 2.a: Consider $\tilde{\mu} = \tilde{\mu}_{c}^{(2)}$, defined
by Eq.\ (\ref{eqA18}). This
situation is possible when $0 \leq \tilde{\alpha} \leq 1.054444$.
Within this range of values for $\tilde{\alpha}$, 
the parameter $\frac{a_1^3}{27}+\frac{a_2^2}{4}\ge 0$. Then, we find
a single real solution

\begin{eqnarray}
\xi_{c} =  \frac{1}{39}\left(7-18\tilde{\alpha}+
\sqrt{49-18\tilde{\alpha}-27\tilde{\alpha}^2} \right)+s_1 + s_2
\label{eqA22}
\end{eqnarray}

Case 2.b: Consider $\tilde{\mu} = \tilde{\mu}_{c}^{(3)}$, defined by
Eq.\ (\ref{eqA18b}). This situation
is possible when $1 \leq \tilde{\alpha} \leq 1.054444$. 
Within this range of values for $\tilde{\alpha}$,
the parameter $\frac{a_1^3}{27}+\frac{a_2^2}{4}\ge 0$. Then, we find
a single real solution

\begin{eqnarray}
\xi_{c} =  \frac{1}{39}\left(7-18\tilde{\alpha}-
\sqrt{49-18\tilde{\alpha}-27\tilde{\alpha}^2} \right)+s_1 + s_2
\label{eqA22b}
\end{eqnarray}

Case 3: $a_3 \ne 0$, $a_2 \ne 0$. In this case, we consider again
the general quartic Eq. (\ref{eqA8}). Following Ferrari's method
\cite{Abramowitz}, we find 4 possible roots
\begin{eqnarray}
\xi_{c,(1,2)} = -\frac{B}{4A} + \frac{1}{2}P \pm \frac{1}{2}Q\nonumber\\
\xi_{c,(3,4)} = -\frac{B}{4A} - \frac{1}{2}P \pm \frac{1}{2}U
\label{eqA23}
\end{eqnarray}
Here, we defined
\begin{eqnarray}
P = \sqrt{\frac{B^2}{4 A^2}-\frac{C}{A}+y_{1}}
\label{eqA24}
\end{eqnarray}
\begin{eqnarray}
Q = \left\{\begin{array}{cc}\sqrt{\frac{3}{4}\frac{B^2}{A^2}-P^2
-2\frac{C}{A}+\frac{1}{4}(4\frac{BC}{A^2}-8\frac{D}{A}
-\frac{B^3}{A^3})P^{-1}}, 
& P\neq 0\\
\sqrt{\frac{3}{4}\frac{B^2}{A^2}-2\frac{C}{A}+2\sqrt{y_{1}^{2}
-4\frac{E}{A}}}, & P=0\end{array} \right.
\label{eqA25}
\end{eqnarray}
\begin{eqnarray}
U=\left\{\begin{array}{cc}\sqrt{\frac{3}{4}\frac{B^2}{A^2}-P^2 - 2\frac{C}{A}
-\frac{1}{4}(4\frac{BC}{A^2}-8\frac{D}{A}-\frac{B^3}{A^3})P^{-1}},
& P \neq 0\\
\sqrt{\frac{3}{4}\frac{B^2}{A^2}-2\frac{C}{A}-2\sqrt{y_{1}^2-4\frac{E}{A}}},
& P = 0\end{array} \right.
\label{eqA26}
\end{eqnarray}
From Eq.\ (\ref{eqA23}), the largest real root corresponds to the
physical solution of Eq.\ (\ref{eq79}).

The parameter $y_{1}$ in Eqs.\ (\ref{eqA24}--\ref{eqA26})
is obtained as the real root of the auxiliary cubic equation
\begin{eqnarray}
y^{3} + \gamma_{2}y^2 + \gamma_{1} y + \gamma_{0} = 0
\label{eqA26a}
\end{eqnarray}
Here, we defined the parameters
\begin{eqnarray}
\gamma_{2} &=& -\frac{C}{A}\nonumber\\
\gamma_{1} &=& \frac{BD}{A^2}-4\frac{E}{A}\nonumber\\
\gamma_{0} &=& 4\frac{CE}{A^2}-\frac{D^2}{A^2}-\frac{B^2 E}{A^3}
\label{eqA26b}
\end{eqnarray}
Let us define
\begin{eqnarray}
q &=& \frac{\gamma_{1}}{3}-\frac{\gamma_{2}^2}{9}\nonumber\\
r &=& \frac{1}{6}(\gamma_{1}\gamma_{2}-3\gamma_{0})-\frac{\gamma_{2}^{3}}{27}
\nonumber\\
\Delta &=& q^3 + r^2
\label{eqA26c}
\end{eqnarray}
We have three possible cases: $\Delta>0$, $\Delta=0$, and $\Delta<0$.

Case 3.a: $\Delta > 0$. In this case, we have one real root $y_{1}$
for the auxiliary cubic Eq.\ (\ref{eqA26a}),
and two complex roots. The real root to be used in 
Eqs.\ (\ref{eqA23}--\ref{eqA26}) is given by
\begin{eqnarray}
y_{1} = (r + \Delta^{1/2})^{1/3} +  (r - \Delta^{1/2})^{1/3}
- \frac{\gamma_{2}}{3}
\label{eqA26d}
\end{eqnarray}

Case 3.b: $\Delta=0$. In this case, all roots of the
auxiliary cubic Eq.\ (\ref{eqA26a}) are real, with two of
them identical, and given by
\begin{eqnarray}
y_{1} &=& 2 r^{1/3} - \frac{\gamma_{2}}{3}\nonumber\\
y_{2} &=& y_{3} = - r^{1/3}
\label{eqA26e}
\end{eqnarray} 
In this case, we take the root $y_{1}$ in Eq. (\ref{eqA26e}), to be used
in the formulas Eqs.\ (\ref{eqA23}--\ref{eqA26}).

Case 3.c: $\Delta < 0$. In this case, all three roots 
of the auxiliary cubic Eq.\ (\ref{eqA26a}) are real and
different.
\begin{eqnarray}
y_{1} &=& 2(r^2 -\Delta)^{1/6}\cos(\theta/3)-\frac{\gamma_{2}}{3}\nonumber\\
y_{2} &=& -2(r^2 -\Delta)^{1/6}\cos(\theta/3+\pi/3)
-\frac{\gamma_{2}}{3}\nonumber\\
y_{3} &=& -2(r^2 -\Delta)^{1/6}\cos(\theta/3-\pi/3)
-\frac{\gamma_{2}}{3}
\label{eqA26f}
\end{eqnarray}
Here,
$\theta = \tan^{-1}\left(\frac{(-\Delta)^{1/2}}{r}\right)$. We take
the root $y_{1}$ to be used in Eqs.\ (\ref{eqA23}--\ref{eqA26}).

\section*{Appendix 3}

The same arguments based on the homogeneous property of $\lambda_{\rm{max}}$
and application of Euler's theorem can be repeated for the case of the
Eigen model, to obtain a solution for the non-symmetric case,
starting from Eq.\ (\ref{eq108b}).
Now, the matrix to consider is
\begin{eqnarray}
{\bf{M}}(\{\bar{x}_{c}^{\alpha}\},\{\bar{\eta}^{\alpha\beta}\})
=\sum_{\alpha\ne\beta=1}^{h}\bar{\eta}_{c}^{\alpha\beta}\tau^{\alpha\beta}
+ \sum_{\alpha=1}^{h-1}\bar{x}_{c}^{\alpha}\Theta^{\alpha}
\label{eqA5_1}
\end{eqnarray}
Since ${\bf{M}}$ is clearly a homogeneous function of degree $k=1$ in the
vector $(\bar{x}_{c}^{1},\bar{x}_{c}^{2},\ldots,\bar{x}_{c}^{h-1},
\{\eta^{\alpha\beta}_{c}\})$,
$\lambda_{{\rm{max}}}$ is homogeneous of degree $k=1$ as well 
(the proof is identical as in Property I, Appendix 1 for the parallel model).  
Therefore, after Euler's theorem (see Appendix 1), we have
the identity
\begin{eqnarray}
\lambda_{{\rm{max}}} -\sum_{\alpha=1}^{h-1}
\bar{x}^{\alpha}_{c} x^{\alpha}_{c} 
-\sum_{\alpha\ne\beta=1}^{h}\bar{\eta}_{c}^{\alpha\beta}
\eta_{c}^{\alpha\beta}
=0
\label{eqA5_2}
\end{eqnarray}
After the saddle-point Eqs.\ (\ref{eq107})--(\ref{eq108}), we obtain
the same components for the eigenvalue $\vec{y}_{\rm{max}}$ as in
the parallel case, Eq.\ (\ref{eqA2_24}),
\begin{eqnarray}
y^{\alpha} = \left\{\begin{array}{cc}\sqrt{x^{\alpha}_{c}}, &
1\leq\alpha\leq h-1\\\sqrt{1 - \sum_{\gamma=1}^{h-1}x^{\gamma}_{c}}, &
\alpha = h\end{array}\right.
\label{eqA5_3}
\end{eqnarray}

Thus, by applying Property III (Appendix 1) for the Eigen model,
we obtain
\begin{eqnarray}
\eta_{c}^{\alpha\beta}&=&\langle \tau^{\alpha\beta} \rangle=
\vec{y}_{\rm{max}}^{T}\tau^{\alpha\beta}\vec{y}_{\rm{max}}
\nonumber\\
&=&  
\left\{\begin{array}{cc}
\sqrt{x^{\alpha}_{c}x^{\beta}_{c}} 
+ 1 - x^{\alpha}_{c}, & \alpha\ne\beta\ne h\\
\sqrt{x^{\alpha}_{c}\left(1 - \sum_{\gamma=1}^{h-1}x^{\gamma}_{c}\right)}
+ 1 - x^{\alpha}_{c}, & \beta = h,\, \alpha\ne h\\
\sqrt{\left(1-\sum_{\gamma=1}^{h-1}x^{\gamma}_{c}\right)x^{\beta}_{c}}
+ \sum_{\gamma=1}^{h-1}x^{\gamma}_{c}, & \alpha = h,\,\beta\ne h
\end{array}\right.
\label{eqA5_4}
\end{eqnarray}

Thus, the mean fitness for the $h$-states Eigen model under arbitrary
mutation scheme is given by the expression
\begin{eqnarray}
f_{m} &=& \max_{\{x^{1}_{c},x^{2}_{c},\ldots,x^{h-1}_{c}\}}\bigg\{
f(x^{1}_{c},x^{2}_{c},\ldots,x^{h-1}_{c}) e^{\sum_{\alpha\ne\beta=1}^{ h-1}
\mu_{\alpha\beta}[\sqrt{x^{\alpha}_{c}x^{\beta}_{c}} - x^{\alpha}_{c}]}
\nonumber \\ && \times
e^{\sum_{\alpha=1}^{h-1}\mu_{\alpha h}[\sqrt{x^{\alpha}_{c}\left(
1-\sum_{\gamma=1}^{h-1}x^{\gamma}_{c}\right)}-x^{\alpha}_{c}]
+
\sum_{\beta=1}^{h-1}\mu_{h\beta}
[\sqrt{\left(1-\sum_{\gamma=1}^{h-1} x^{\gamma}_{c}\right)x^{\beta}_{c}}
+\sum_{\gamma = 1}^{h-1}x^{\gamma}_{c} - 1]} 
\nonumber \\ && 
- d(x^{1}_{c},x^{2}_{c},\ldots,x^{h-1}_{c})
\bigg\}
\label{eqA5_5}
\end{eqnarray}

\bibliography{boson4}
\end{document}